\documentclass[journal,letter,10pt,twocolumn]{IEEEtran}
\usepackage{amsmath,graphicx}
\usepackage{bbm}
\usepackage{amsmath, amssymb, amscd, amsthm, amsfonts}
\usepackage{subfiles}
\usepackage{color}

\usepackage[caption=false,font=footnotesize]{subfig}
\pdfoutput=1

\usepackage[backend=biber, style=ieee, dashed=false,isbn=false,doi=false,url = false,bibencoding=utf8,citestyle=numeric-comp]{biblatex}
\newtheorem{theorem}{Theorem}

\newtheorem{proposition}{Proposition}

\title{Short-time Fourier Transform-based Signal Recovery for Modulo Analog-to-Digital Converters}

\author{
   \IEEEauthorblockN{Neil Irwin Bernardo, \textit{Member, IEEE}
   }
    \thanks{N.I. Bernardo is with the Electrical and Electronics Engineering Institute, University of the Philippines Diliman, Quezon City 1101, Philippines (e-mail: neil.bernardo@eee.upd.edu.ph).\\
    This author acknowledges the Office of the Chancellor of the University of the Philippines Diliman, through the Office of the Vice Chancellor for Research and Development, for funding support through the PhD Incentive Award Grant 252509 YEAR 1.}
   }
\begin{document}
%\ninept
%
\maketitle
\begin{abstract}
This study introduces a short-time Fourier transform–based method for reconstructing signals encoded using modulo analog-to-digital converters with 1-bit folding information. In contrast to existing Fourier-based reconstruction approaches that require complete access to the entire observation, the proposed technique performs reconstruction over short, overlapping segments, enabling significantly lower latency while preserving the recovery accuracy. We also address the spectral leakage introduced by the windowing operation by selecting window parameters that balance the leakage suppression and the computational complexity of the algorithm. In addition, we establish conditions under which the correct unfolding of the modulo samples is guaranteed, leading to a reconstruction error determined solely by the quantization noise at the output. The numerical results demonstrate that the proposed method enables modulo analog-to-digital converters to surpass the mean squared error performance of conventional analog-to-digital converters. Furthermore, the proposed recovery method offers improved reconstruction performance compared with higher-order difference–based recovery, particularly in low-resolution and low–sampling rate regimes.
\end{abstract}
\section{Introduction}
\label{section:intro}

An analog-to-digital converter (ADC) is a critical element of a data acquisition system because it is responsible for the transformation of continuous-time analog observations from the physical world into \textcolor{black}{a} format that can be further processed or stored in digital signal processing (DSP) pipelines. The conversion process in conventional ADCs involves three operations: (1) a \emph{sampler} that acquires the amplitude of the continuous-time input at uniform time intervals; (2) a \emph{quantizer} that maps the amplitude samples into a finite set of discrete values; and (3) an \emph{encoder} that converts the quantized values into binary representations \cite{Eldar}. The sampling rate and quantizer resolution of the ADC were chosen according to a fidelity criterion and a power consumption budget. Theoretically, the sampling rate requirement for alias-free reconstruction and the quantization resolution needed to achieve some distortion level are governed by the Shannon-Nyquist sampling theorem \cite{Shannon:1949} and the rate-distortion theory \cite{berger1971rate}, respectively. However, increasing these parameters also incurs a substantial increase in the ADC power consumption. Specifically, the ADC power consumption scales exponentially with the number of quantization bits and linearly with the sampling rate \cite{Walden:1999}.

Another important parameter that affects the performance of an ADC is its dynamic range (DR). Owing to the finite number of quantization bits in an ADC, a trade-off exists between its dynamic range and resolution. An input signal may drive the ADC to saturation if the DR of the input signal exceeds the DR of the ADC. This issue can be mitigated by increasing the DR of the ADC. However, doing so under a fixed quantization bit budget increases the quantization noise power, causing weak input signals to be overwhelmed by the quantization noise. The presence of both weak and strong components in the input signal increases the ADC resolution required to achieve a specific fidelity criterion \cite{Laporte-Fauret:2018}.

%However, increasing the ADC range to accommodate high DR signals would increase the quantization step size. Weak input signals could be buried in the quantization noise. The presence of both weak and strong components in the input signal increases the required ADC resolution to achieve a specific fidelity criterion \cite{Laporte-Fauret:2018}.

A promising solution to address the DR bottleneck of ADCs is the \emph{modulo sampling} framework \cite{Bhandari:2017,Bhandari:2021}. The idea behind modulo sampling is to apply a nonlinear folding operation called \emph{modulo} to the input signal before the ADC. This modulo pre-processing stage ensures that the signal range is confined within the DR of the ADC. Moreover, for a given amplitude quantization bit budget, modulo ADCs can achieve better digital resolution than conventional ADCs as demonstrated in previous works \cite{Zhu:2024b,Guo:2024}.

Because the modulo operation intentionally introduces nonlinear distortion to the signal, several studies on modulo sampling have focused on the development of robust recovery techniques to correct the distortion induced by the modulo operation. Reconstruction techniques for modulo sampling include higher-order differences-based approaches \cite{Bhandari:2021}, discrete Fourier transform (DFT)-based techniques \cite{Fourier-Prony,Azar:2022,Zhang:2023,Shah:2023,Bernardo_ISIT2024,Beckmann:2024}, prediction-based filtering \cite{Romanov:2019,Ordentlich:2018}, iterative methods \cite{Guo:2023}, sparse signal modulo recovery \cite{Bhandari:2018,Musa:2018,Shah:2021,Prasanna:2021}, multi-channel modulo structures \cite{Gan:2020,Gong:2021,Shtendel:2024,Florescu:2025}, and thresholding-based schemes\cite{Florescu:2021,Florescu:2022,Florescu:2022b,Florescu:2022c,Geethu:2025}. There have also been research efforts geared toward the practical hardware implementation of modulo ADCs \cite{Fourier-Prony,krishna,Fernandez-Menduina:2022,Mulleti:2023,Zhu:2024,Shah:2024} and their potential applications in radar systems, communication systems, imaging, and electroencephalogram (EEG) recovery \cite{Bhandari:2020,Ordonez:2021,Feuillen:2022, Feuillen:2023, Zhang:2023, Shtendel:2023, Geng:2023,Liu:2023, Shtendel:2024, Liu:2025, Zhang:2025, Mulleti:2025, Yan:2025}.

Another important direction in modulo ADC research is the establishment of theoretical performance guarantees to reconstruct the unfolded signal from the modulo samples. In \cite{Bhandari:2021}, it was shown that the higher-order differences-based reconstruction approach can perfectly reconstruct a finite-energy bandlimited (BL) signal from its modulo samples if the sampling rate used is at least $2\pi e$ ($\approx 17.07$) times the Nyquist rate. This sufficient condition for the sampling rate significantly increases in the presence of bounded noise \cite[Theorem 3]{Bhandari:2021}. For finite-energy BL signals, perfect reconstruction is possible for any sampling rate approaching the Nyquist rate from above by using a prediction filter with a sufficiently long filter length \cite{Romanov:2019}. This approach does not require any side information such as knowledge of unfolded samples and the auto-correlation function. However, the proposed prediction filter-based approach does not take into account the impact of finite quantization. Theoretical guarantees for the Fourier-based reconstruction were established in \cite{Fourier-Prony}. However, this approach suffers from spectral leakage and may require long observation windows. To avoid spectral leakage, recovery methods based on time-domain thresholding can be used \cite{Florescu:2021,Florescu:2022,Florescu:2022b,Florescu:2022c,Geethu:2025}. Such methods work for generalized modulo sampling models that consider modulo hysteresis and folding transients. However, the recovery guarantees established for thresholding-based recovery methods fail under ideal modulo nonlinearities\textcolor{black}{, as explained} in Section \ref{subsection:compare_recovery}. The recently developed UNO framework \cite{UNO} integrates the Unlimited Sampling Framework (USF) \cite{Bhandari:2021} with 1-bit quantization, where each folded sample is processed by multiple 1-bit quantizers using varying thresholds. In effect, the set of 1-bit quantizers operates collectively as a multi-bit quantizer. Furthermore, the accompanying theoretical guarantees are expressed through high-probability bounds.

A recent line of research on modulo ADCs with 1-bit folding side information \cite{Bernardo_ISIT2024, Bernardo2025TSP} established a mean squared error (MSE) performance guarantee when the oversampling factor (OF) exceeds three and the quantizer resolution exceeds three bits. The 1-bit side information identifies the time indices at which folding occurs. However, the analysis in these studies neglects both spectral leakage effects and the contribution of quantization noise at certain time indices. When the 1-bit folding side information is unavailable, the same MSE guarantees can be recovered using an orthogonal matching pursuit (OMP)–based reconstruction scheme \cite{Bernardo2025TSP}, although this requires a stricter condition on the amplitude quantization bit budget. Additionally, the Fourier-based recovery methods in \cite{Bernardo_ISIT2024, Bernardo2025TSP} incur very high computational complexity, rendering them impractical for very long sequences.

In this study, we propose a short-time Fourier transform (STFT)–based recovery method for reconstructing the original input samples from the output of a modulo ADC equipped with 1-bit folding side information. The method performs unfolding on short, overlapping frames of the signal, enabling successive, low-latency reconstruction rather than requiring access to the entire sequence at once. We employ a smooth window function with controlled roll-off factor to mitigate the spectral leakage introduced by the STFT windowing operation. We further established a mean-squared error (MSE) performance guarantee for the proposed method and validated the analysis through numerical experiments. This STFT-based recovery framework substantially reduces the computational complexity relative to the Fourier-based recovery schemes, especially for long sequences. Moreover, we provide a more comprehensive analysis of the MSE guarantees by considering the contribution of the spectral leakage effects inherent in Fourier-based recovery methods. This has not been accounted for in the earlier works.

\begin{figure*}[t!]
    \centering
    \includegraphics[scale = .925]{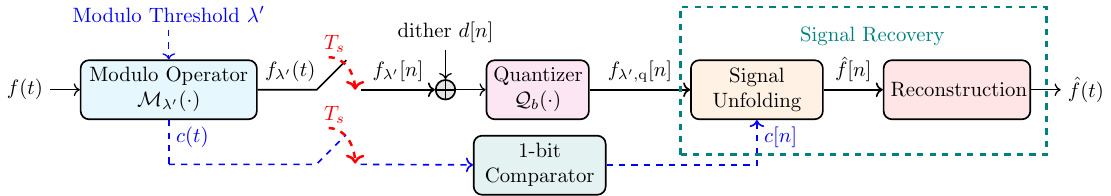}
    \caption{Schematic diagram of the modulo ADC with 1-bit folding information. \textcolor{black}{The input signal is wrapped by the modulo operation whenever it exceeds the converter’s dynamic range, producing a folded output sequence. A folding detector simultaneously generates a continuous-time folding signal $c(t)$, and the corresponding discrete-time folding information $c[n]$ is obtained by sampling this signal.}}
    %The 1-bit folding information $c[n]$ is obtained by sampling $c(t)$.}
    \label{fig:modulo_ADC}
\end{figure*}

\begin{figure}[!t]
    \centering
    \includegraphics[scale = .925]{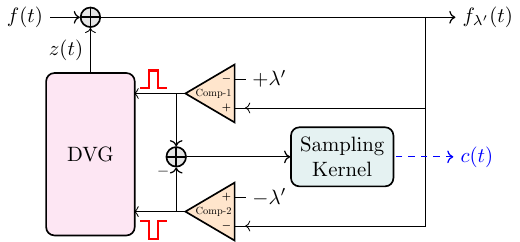}
    \caption{\textcolor{black}{Modulo operator mechanism to generate the folded signal $f_{\lambda'}(t)$ and the folding information signal $c(t)$. The comparators detect folding events and provide control signals to the discrete voltage generator that produces the wrapped output. At the same time, comparator outputs are processed to generate the continuous-time folding information $c(t)$.} }
    %\caption{Folding mechanism and folding information generation of the modulo operator block. }
    \label{fig:modulo_operator}
\end{figure}

The main contributions of this work are summarized as follows:
\begin{itemize}
    \item We develop an STFT-based recovery algorithm to reconstruct the input signal from the output of a modulo ADC equipped with 1-bit folding information. Unlike existing Fourier-based reconstruction techniques, the proposed method does not require the processing of a long observation window or the entire data sequence before producing unfolded samples, thereby substantially reducing the computational complexity. Moreover, the spectral leakage introduced by short-frame processing is mitigated through the use of a smoother window function with gradual transitions at the boundaries. The STFT window design considerations and computational complexity of the proposed algorithm are analyzed in Sections \ref{subsection:window_design} and \ref{subsection:complexity}, respectively.
    \item We derive sufficient conditions for the oversampling factor ($\mathrm{OF}$) and quantizer resolution $\mathrm{b}$ under which the proposed STFT-based algorithm reliably unfolds the modulo ADC output. When these conditions are met, the only source of the reconstruction error is the quantization noise. The MSE performance guarantee is formalized in Theorem \ref{theorem:MSE_guarantee}.
    \item We compare the MSE of a modulo ADC equipped with the proposed STFT-based recovery algorithm against that of (i) a conventional ADC without modulo operation and (ii) a modulo ADC using the higher-order differences (HoD)-based recovery method. Under the established sufficient conditions, we show that the MSE of the STFT-based recovery scales as $\mathcal{O}\!\left(1/\mathrm{OF}^{3}\right)$ when spectral leakage is negligible, and as $\mathcal{O}\!\left(1/\mathrm{OF}^{2}\right)$ when leakage is present. Here, $\mathrm{OF}$ denotes the oversampling factor \textcolor{black}{as} defined in Section~\ref{section:sys_model}. In contrast, the MSE of a conventional ADC scales only as $\mathcal{O}\!\left(1/\mathrm{OF}\right)$. We further observed that the HoD-based recovery method fails to reliably unfold the modulo ADC output at certain oversampling factors when $b = 4$ or $5$. Numerical results are provided to illustrate these behaviors and validate the theoretical predictions.
\end{itemize}

The remainder of the paper is organized as follows: Section II describes the system model and the proposed reconstruction algorithm. Section III presents the theoretical guarantees of the proposed recovery algorithm in terms of recovery performance and computational complexity. Section IV provides the numerical results that validates the theoretical analysis of the proposed method. Section V concludes the paper.

\emph{Notation:} The following notations are used throughout this paper. The sets of real numbers, integers, and natural numbers are denoted by $\mathbb{R}$, $\mathbb{Z}$, and $\mathbb{N}$, respectively. When referring to discrete-time signals, the notation $z[n]$ is used to signify $z(nT_{\mathrm{s}})$, assuming that the sampling period $T_{\mathrm{s}}$ is evident in the context. The first-order difference of a discrete-time signal $x[n]$ is denoted by $\underline{x}[n] = x[n] - x[n-1]$, where $x[-1] = 0$ unless stated otherwise. Vectors and matrices are written in bold format (e.g., $\mathbf{z}$, $\mathbf{A}$) while sets are written in calligraphic format (e.g., $\mathcal{S}$). The cardinality of a set $\mathcal{S}$ is written as $|\mathcal{S}|$. If $\mathbf{x}$ is a vectorized form of some discrete-time signal $x[n]$, then $\underline{\mathbf{x}}$ is the vectorized form of $\underline{x}[n]$. The $\ell_p$-norm of a vector $\mathbf{x}$ is written as $\|\mathbf{x}\|_{p}$. The set of values of the discrete-time signal $x[n]$ for $n\in\mathcal{S}$ is denoted as $x_{\mathcal{S}}[n]$. The Moore-Penrose inverse of the matrix $\mathbf{A} \in \mathbb{R}^{m \times n}$ is denoted by $\mathbf{A}^{\dagger}$. We use $\omega$ and $\Omega$ to denote the frequencies for the continuous-time Fourier Transform (CTFT) and discrete-time Fourier Transform (DTFT) spectra, respectively. We also use the standard Big-O notation  $\mathcal{O}(\cdot)$ to describe the asymptotic growth rates in this paper.

\section{System Model and Reconstruction Algorithm}
\label{section:sys_model}

In this section, we describe the signal acquisition process of the modulo ADC and introduce the proposed short-time Fourier Transform (STFT)-based recovery framework. We first present the hardware model and mathematical formulation of the modulo folding operation, including the generation of the 1-bit folding information that indicates the folding instances. We also outline the proposed STFT-based signal recovery scheme and highlight its advantages over existing Fourier-based recovery methods. We also \textcolor{black}{provide} design guidelines for selecting the STFT window parameters.

\subsection{Signal Acquisition Process}

We consider the modulo ADC system shown in Figure \ref{fig:modulo_ADC}. The input $f(t)$ is a bandlimited signal with (angular) frequency support $\left[-\frac{\omega_m}{2},\;+\frac{\omega_m}{2}\right]$. This signal is folded by the modulo operator block to produce the modulo signal $f_{\lambda'}(t)$. The operation behind the modulo folding mechanism is illustrated in Figure \ref{fig:modulo_operator}. This feedback structure for the folding operation is adopted from \cite{Mulleti:2023}. Comparator 1 triggers a positive \textcolor{black}{output} whenever $f_{\lambda'}(t)$ crosses $+\lambda$ from below. Similarly, comparator 2 triggers a negative \textcolor{black}{output} whenever $f_{\lambda'}(t)$ crosses $-\lambda$ from above. These trigger signals are fed to a discrete voltage generator (DVG). The voltage level of the DVG is increased (resp. decreased) by $2\lambda'$ whenever it receives a positive (resp. negative) trigger signal. A detailed circuit-level implementation of the DVG is presented in \cite{Mulleti:2023}. Mathematically, the folded signal can be expressed as
\begin{align}
    f_{\lambda'}(t) = \bigg[\left(f(t)+\lambda'\right)\;\text{mod}\; 2\lambda'\bigg] -\lambda',
\end{align}
where $\lambda'\in (0,\|f(t)\|_{\infty})$ is the modulo threshold. The $\ell_{\infty}$-norm of a waveform is its maximum amplitude. Throughout this paper, we assume ideal modulo nonlinearities. That is, there are no modulo hysteresis and folding transients like in the generalized modulo sampling model presented in \cite{Florescu:2022}. Furthermore, we also assume that the amplitude of the first sample of $f(t)$, i.e. $f[0]$, is within $[-\lambda',\lambda']$. This ensures that no unknown constant offset $2\lambda'p$, where $p\in\mathbb{Z}$, is present in the reconstructed signal.

To digitize the modulo signal, $f_{\lambda'}(t)$ is first sampled at every $T_{s}$ seconds. The angular sampling rate is $\omega_s = \frac{2\pi}{T_{s}} = \mathrm{OF}\times \omega_{m}$, where $\mathrm{OF} \geq 1$ is the oversampling factor. We also define $\rho = \frac{1}{\mathrm{OF}}$. Prior to quantization, a (non-subtractive) dither sequence $d[n]$ is added to $f_{\lambda'}[n]$. Samples of the dither sequence are drawn i.i.d. from a triangle distribution with amplitude support $\big(-\frac{2\lambda}{2^b},+\frac{2\lambda}{2^b}\big]$. The rationale for using triangle dither in the modulo ADC problem setup is explained in Section \ref{subsection:stat_quant_noise}. The resulting signal is then forwarded to the $b$-bit uniform scalar quantizer $\mathcal{Q}_b(\cdot)$ with range $[-\lambda,+\lambda]$. Consequently, each quantization bin has width $\frac{2\lambda}{2^b}$. To prevent the quantizer from being overloaded, the quantizer DR is set to $\lambda = \frac{2^b \lambda'}{2^b - 2}$.

The quantizer output can be written as
\begin{align}
    f_{\lambda',\mathrm{q}}[n] =& \mathcal{Q}_b(f_{\lambda'}[n] + d[n])\nonumber\\
    =& f[n] + z[n] + \epsilon[n],
\end{align}
where $z[n] = f_{\lambda'}[n] - f[n]\in 2\lambda'\mathbb{Z}$ is the residual samples due to the folding operation and $\epsilon[n] = \mathcal{Q}_b(f_{\lambda'}[n] + d[n]) - f_{\lambda'}[n]$ is the quantization noise sequence.

\begin{figure*}[!t]
    \centering
    \includegraphics[width = .95\textwidth]{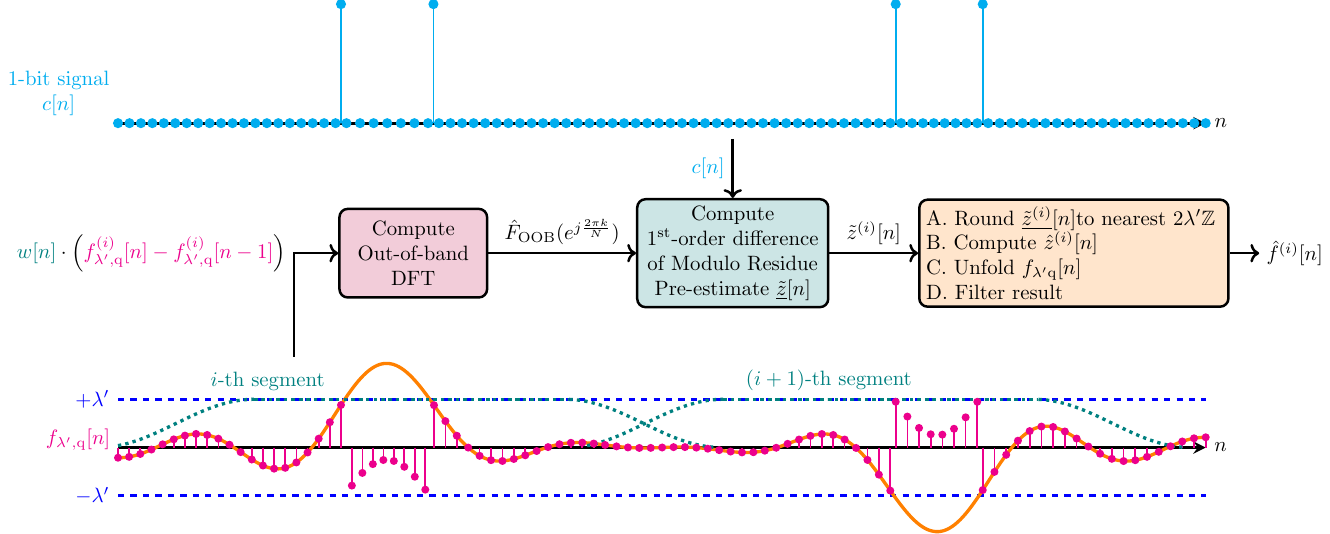}
    \caption{Overview of the proposed STFT-based recovery method for modulo sampling. The proposed recovery method is composed of three steps: (1) out-of-band DFT computation, (2) estimation of the modulo residue via 1-bit folding information and out-of-band DFT values, and (3) removal of modulo residue from folded signal.} 
    \label{fig:algorithm_visual}
\end{figure*}

In addition to $f_{\lambda'}(t)$, the modulo operator also generates a signal $c(t)$ as shown in Figure \ref{fig:modulo_operator}. This signal is a finite rate of innovation (FRI) signal and can be expressed as
\begin{align}
    c(t) =& g_2(t)\ast \left(\sum_{k}g_1(t-\tau_{k})\right)\nonumber\\
    =& \sum_{k}g(t-\tau_{k}),
\end{align}
where $\tau_k$ is the location of the level crossings, $g_1(t)$ is the pulse shape of the trigger signals, $g_2(t)$ is the sampling kernel, and $g(t) = (g_2\ast g_1)(t)$. To simplify the analysis, $g(t)$ is assumed to be a rectangular pulse with width $T_{s}$ and unity amplitude. The signal $c(t)$ is sampled at rate $\omega_s$ and then fed to a 1-bit comparator to produce $c[n]$. In essence, $c[n]$ provides information on whether the signal \textcolor{black}{crosses} either $+\lambda'$ or $-\lambda'$ in the interval $(nT_{\mathrm{s}},(n+1)T_{\mathrm{s}}]$. We note that even though our modulo ADC generates two channel streams, our setup is different from the multi-channel modulo sampling structures presented in \cite{Gan:2020,Gong:2021,Shtendel:2024}. The multiple channels in \cite{Gan:2020, Gong:2021,Shtendel:2024} use different modulo thresholds to process the same signal, thereby generating two multi-bit modulo ADC output sample streams. In contrast, our second stream only indicates the position of the folding instances in the first stream. The information carried by this second stream can be represented as a single bit per sample.

\subsection{Is the 1-bit Folding Information Practical?}
One might wonder whether generating 1-bit folding information $c[n]$ is practical when implementing a modulo ADC. Early works on modulo ADC hardware considered recovery methods that utilize the reset count in addition to the self-reset ADC output samples \cite{Rhee:2003,Sasagawa:2016}. However, implementing a self-reset ADC with a multi-bit reset count signal may incur a significant increase in power consumption and circuit complexity, especially when the modulo threshold is small \cite{Bhandari:2021}. Furthermore, having a multi-bit reset count signal with the same resolution as the self-reset ADC output samples \textcolor{black}{would double} the storage requirements of the system. Limiting $c[n]$ to 1-bit solves these problems.

As shown in Figures \ref{fig:modulo_ADC} and \ref{fig:modulo_operator}, generating $c[n]$ requires only one additional comparator and one adder from the ``no $c[n]$'' modulo ADC setup. The additional circuitry has minimal impact on the overall power consumption, and its penalty is fixed regardless of $b$. For instance, consider a flash ADC architecture for amplitude quantization. An 8-bit modulo ADC with no $c[n]$ would have $2^8 + 2 = 258$ comparators\footnote{The 2 additional comparators come from the comparators of the folding mechanism.}, whereas an 8-bit modulo ADC with 1-bit folding information $c[n]$ would have 259 comparators. The difference in the comparator counts is fixed regardless of the number of bits for amplitude quantization. We also note that the folding information generated in Figure \ref{fig:modulo_ADC} is one bit less than that generated by the modulo ADC hardware proposed in \cite{krishna}. In \cite{krishna}, the 2-bit signal indicates three states: `no crossing', `crossed $+\lambda'$ threshold', `crossed $-\lambda'$ threshold'.

Finally, while the above discussion emphasizes low circuit complexity, it is important to note that our recovery algorithm assumes an ideal timing alignment between $c[n]$ and $f_{\lambda',\mathrm{q}}[n]$. In practical implementations, latency, jitter, or other synchronization imperfections can affect the recovery performance. Evaluating the robustness of the STFT-based recovery to such timing offsets is an important topic for future research. Nevertheless, we note that a modulo ADC with 1-bit folding information has already been demonstrated in hardware \cite{Shah:2024}, where both the modulo ADC output and 1-bit folding information are generated using a single clock. This design choice inherently mitigates the timing alignment issues. The specific architecture in \cite{Shah:2024} differs from that depicted in Figures \ref{fig:modulo_ADC} and \ref{fig:modulo_operator}, but it provides a concrete example showing that generating a synchronized 1-bit folding signal is feasible in practice.

\subsection{Signal Recovery Scheme}
\label{subsection:signal_recovery}

The proposed recovery algorithm first unfolds $f_{\lambda',\mathrm{q}}[n]$ by estimating $z[n]$ and then reconstructs the continuous-time signal from the result. Since the sampling frequency used is above the Nyquist rate, the continuous-time signal can be uniquely identified using an appropriate reconstruction filter. In this subsection, we discuss the unfolding mechanism. An overview of the proposed unfolding procedure is shown in Figure \ref{fig:algorithm_visual}. The unfolding scheme multiplies a window \textcolor{black}{function} $w[n]$ to \textcolor{black}{the} small overlapping segments of $f_{\lambda',\mathrm{q}}[n]$ and then recovers the modulo residue signal at each segment. The window function is applied sequentially across the signal with a fixed hop size of $N(1-\frac{\alpha}{2})$ until the entire $f_{\lambda',\mathrm{q}}[n]$ is processed.

%The window is slid from left to right until the whole $f_{\lambda',\mathrm{q}}[n]$ is processed.

Let $N$ be the number of samples in a short segment of $f_{\lambda',\mathrm{q}}[n]$. A length-$N$ tapered cosine window $w[n]$ (also known as Tukey window \cite{MathWorksTukeyWindow}) can be written as
\begin{align}
    w[n] = \begin{cases}
        \frac{1+\cos\left(\frac{2\pi}{\alpha}\left(\frac{n}{N} - \frac{\alpha}{2}\right)\right)}{2}\;\;\;,\; 0\leq n< \frac{\alpha N}{2}\\
    \qquad\qquad1\;\quad\qquad,\; \frac{\alpha N}{2} \leq n < N\left(1-\frac{\alpha}{2}\right)\\
        \frac{1+\cos\left(\frac{2\pi}{\alpha}\left(\frac{n}{N} - 1 +  \frac{\alpha}{2}\right)\right)}{2},\; N\left(1-\frac{\alpha}{2}\right) \leq n < N,
    \end{cases}
\end{align}
where the parameter $\alpha > 0$ denotes the roll-off factor. We select $\alpha$ such that $\alpha N \in 2\mathbb{Z}$. 

Suppose that the $i$-th segment of $f_{\lambda',\mathrm{q}}[n]$ is denoted as $f_{\lambda',\mathrm{q}}^{(i)}[n]$ and the $i$-th segment of the unfolded signal as $f^{(i)}[n]$. Consecutive segments overlap such that the last $\frac{\alpha N}{2}$ samples of $f_{\lambda',\mathrm{q}}^{(i-1)}[n]$ are the first $\frac{\alpha N}{2}$ samples $f_{\lambda',\mathrm{q}}^{(i)}[n]$. The window function $w[n]$ is multiplied to the first-order difference of $f_{\lambda',\mathrm{q}}[n]$ to obtain
\begin{align}
    \underline{\tilde{f}_{w}^{(i)}}[n] =& w[n]\cdot \underline{f_{\lambda',\mathrm{q}}^{(i)}}[n]\nonumber\\
    =& w[n]\cdot \left(f_{\lambda',\mathrm{q}}^{(i)}[n] - f_{\lambda',\mathrm{q}}^{(i)}[n-1]\right), 
\end{align}
where $f_{\lambda',\mathrm{q}}^{(i)}[-1] = f_{\lambda',\mathrm{q}}^{(i-1)}[N-1]\;\forall\;i\in\{2,3,\cdots,I\}$ and we assume that $f_{\lambda',\mathrm{q}}^{(1)}[-1] = 0$, i.e. signal starts at zero. We now describe the computation of the modulo residue samples from $f_{\lambda',\mathrm{q}}^{(i)}[n]$. The principle behind the computation of the modulo residue samples is the Fourier domain separation between the bandlimited signal and folding instances \cite{Fourier-Prony}. The length-$N$ Discrete Fourier Transform (DFT) of $\underline{\tilde{f}_{w}^{(i)}}[n]$ is computed as
\begin{align}
    \hat{F}_w(e^{j\frac{2\pi k}{N}}) =& \frac{1}{\sqrt{N}}\sum_{n = 0}^{N-1}\underline{\tilde{f}_{w}^{(i)}}[n]e^{-j\frac{2\pi kn}{N}}.
\end{align}

\begin{figure*}[t!]
    \centering
    \hspace*{-.75cm}
    \includegraphics[width = 1.05\textwidth]{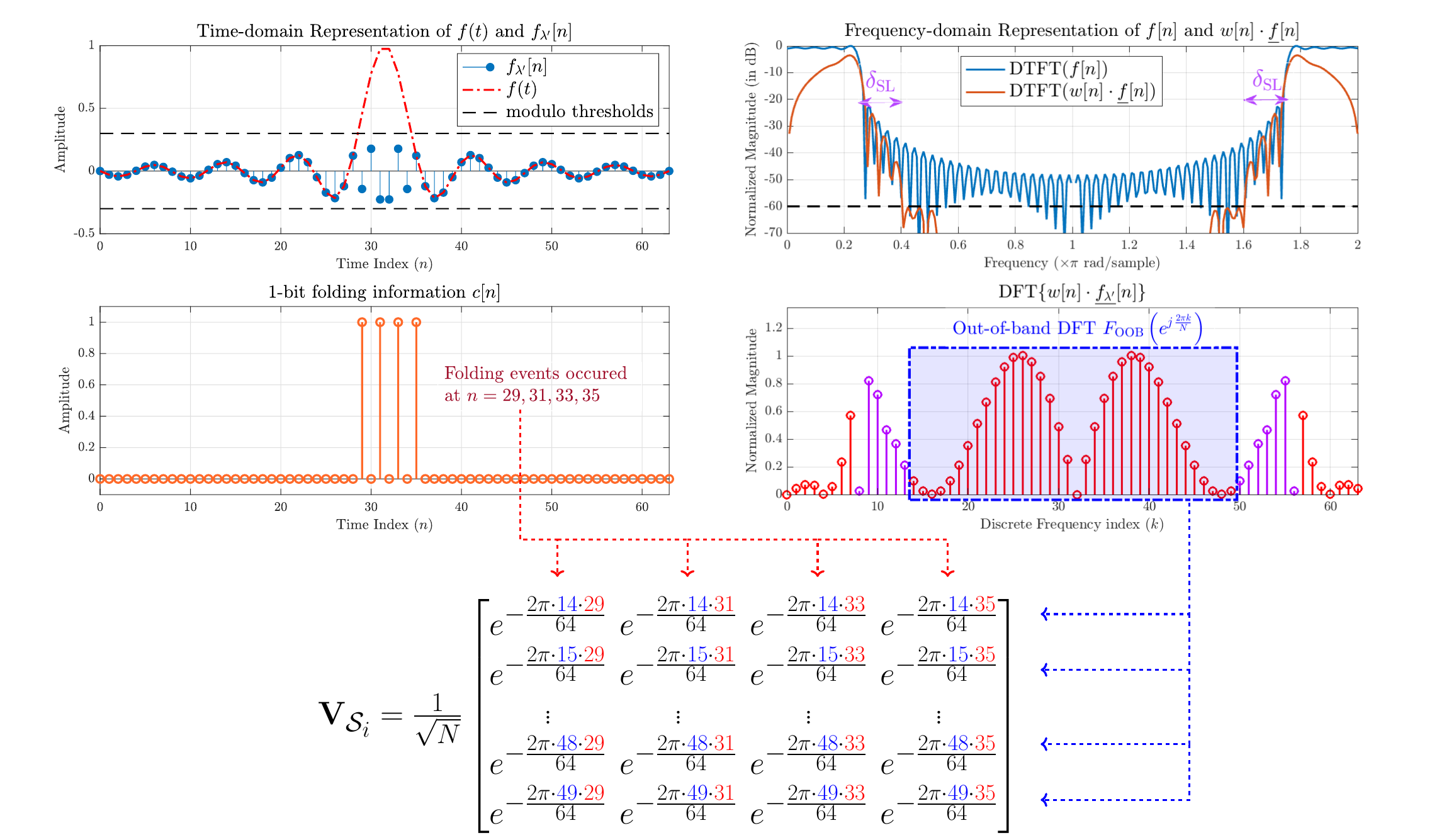}
    \caption{Numerical example illustrating the construction of $\mathbf{V}_{\mathcal{S}_{i}}$ for $N = 64$, $\mathrm{OF} = 4$, and $\alpha = 0.5$. The top-left plot shows the input signal $f(t)$ and the corresponding modulo ADC output $f_{\lambda'}[n]$ when $\lambda' = 0.30$ and no quantization. The folding events are indicated by $c[n]$ in the bottom–left plot. The top–right plot presents the DTFT magnitude spectrum of $f[n] = f(nT_{\mathrm{s}})$ and $w[n]\cdot\underline{f}[n] = w[n]\cdot (f[n] - f[n-1])$. The spectral leakage is visible due to the finite window length. Components below -60 dB are assumed to have negligible impact \textcolor{black}{on} the computation. The length-$\delta_{\mathrm{SL}}$ intervals marked in the top-right plot denote the OOB frequency regions whose leakage exceeds -60 dB. These regions correspond to the $K_{\mathrm{SL}}$ discrete frequency indices marked in the DFT spectrum of the (windowed) 1st-order difference of folded signal $w[n]\cdot \underline{f_{\lambda'}}[n]$, as shown in the bottom–right plot. The matrix $\mathbf{V}_{\mathcal{S}_{i}}$ is constructed by selecting the rows and columns of a DFT matrix corresponding to the OOB discrete frequency indices and folding indices, respectively.
    %The top-left plot shows the output of the modulo ADC when sampling the signal $f(t) = \mathrm{sinc}(t)$ at 4 times the Nyquist rate. The modulo threshold is set to $\lambda' = 0.30$. The folding events are indicated by $c[n]$ in the bottom–left plot. The top–right plot presents the DTFT magnitude spectrum of $f[n] = f(nT_{\mathrm{s}})$ and $w[n]\cdot\underline{f}[n] = w[n]\cdot (f[n] - f[n-1])$, where the spectral leakage is visible due to the finite window length. If spectral components of $\mathrm{DTFT}(w[n]\cdot\underline{f}[n])$ below –60 dB are treated as negligible for the application, then the left and right intervals of length $\delta_{\mathrm{SL}}$ denote the OOB frequency regions whose leakage exceeds -60 dB. These regions correspond to the $K_{\mathrm{SL}}$ discrete frequency indices marked in the DFT spectrum of the (windowed) 1st-order difference of folded signal $w[n]\cdot \underline{f_{\lambda'}}[n]$, as shown in the bottom–right plot. The matrix $\mathbf{V}_{\mathcal{S}_{i}}$ is constructed by selecting the rows and columns of a DFT matrix corresponding to the OOB discrete frequency indices and folding indices, respectively.}
    }
    \label{fig:Vsi_illustration}
\end{figure*}

Due to the finite bandwidth assumption on $f(t)$, $\hat{F}_w(e^{j\frac{2\pi k}{N}})$ for $\frac{2\pi k}{N}\in (\rho\pi+\delta_{\mathrm{SL}}, 2\pi - \rho\pi-\delta_{\mathrm{SL}})$ can be written as
\begingroup
\allowdisplaybreaks
\begin{align}\label{eq:OOB_DFT}
    \hat{F}_{\mathrm{OOB}}(e^{j\frac{2\pi k}{N}}) = &\frac{1}{\sqrt{N}}\sum_{n = 0}^{N-1}w[n]\bigg[\underline{z^{(i)}}[n] + \underline{\epsilon^{(i)}}[n]\bigg]e^{-j\frac{2\pi kn}{N}}\nonumber\\
    =& \frac{1}{\sqrt{N}}\sum_{n \in \mathcal{S}_i}w[n]\bigg[\underline{z_{\mathcal{S}_i}^{(i)}}[n] + \underline{\epsilon_{\mathcal{S}_i}^{(i)}}[n]\bigg] e^{-j\frac{2\pi kn}{N}}\nonumber\\
    &+\frac{1}{\sqrt{N}}\sum_{n \notin \mathcal{S}_i}w[n]\cdot \underline{\epsilon_{\mathcal{S}_i^c}^{(i)}}[n]\cdot e^{-j\frac{2\pi kn}{N}},
\end{align}
\endgroup
where the spectral leakage width, denoted $\delta_{\mathrm{SL}}$, accounts for the frequencies at which spectral leakage is still significant. Some of the energy of $w[n]\cdot\underline{f^{(i)}}[n]$ may leak in the band $(\rho\pi, 2\pi - \rho\pi)$ because of the inherent spectral leakage in the DFT computation and cannot be neglected. Spectral leakage vanishes towards zero at high frequencies and cannot be made zero for any signal. However, the spectral leakage can be controlled by properly choosing the window length $N$ and roll-off parameter $\alpha$. The signals $\underline{z^{(i)}}[n]$ and $\underline{\epsilon^{(i)}}[n]$ are the first-order differences of the modulo residue samples and quantization noise, respectively, in $\tilde{f}_{w}^{(i)}[n]$. The set $\mathcal{S}_i$ contains the indices for which $c[n] = 1$ in the $i$-th segment. \textcolor{black}{This} set identifies the indices in the $i$-th segment for which the DVG changes its output voltage. The signal $z^{(i)}_{\mathcal{S}_i}[n]$ (resp. $z^{(i)}_{\mathcal{S}_i^c}[n]$) corresponds to the modulo residue for all $n$ such that $c[n] = 1$ (resp. $c[n] = 0$). The subscript $\mathrm{OOB}$ in $ \hat{F}_{\mathrm{OOB}}$ indicates that this is the set of DFT coefficients outside the desired signal bandwidth plus non-negligible spectral leakage.

%The sample $z^{(i)}_{\mathcal{S}_i}[n]$ (resp. $z^{(i)}_{\mathcal{S}_i^c}[n]$) correspond to the value of the modulo residue at position $n$ such that $c[n] = 1$ (resp. $c[n] = 0$). The subscript $\mathrm{OOB}$ in $ \hat{F}_{\mathrm{OOB}}$ is used to indicate that this is the set of DFT coefficients outside the desired signal bandwidth plus non-negligible spectral leakage.

We are primarily interested in estimating 
$f^{(i)}[n]$. Our approach is to first estimate $z^{(i)}[n]$ and then subtract $z^{(i)}[n]$ from $f_{\lambda',\mathrm{q}}^{(i)}[n]$. This procedure essentially unfolds the modulo ADC output. Let $K$ be the number of discrete frequencies $k$ such that $\frac{2\pi k}{N}\in (\rho\pi +\delta_{\mathrm{SL}},2\pi - \rho\pi - \delta_{\mathrm{SL}})$. Equation  \eqref{eq:OOB_DFT} can be written in matrix form as
\begin{align}\label{eq:F_OOB}
    \hat{\mathbf{F}}_{\mathrm{OOB}} = \mathbf{V}\left(\underline{\mathbf{\mathbf{z}}_{w}} + \underline{\boldsymbol{\epsilon}_{w}}\right),
\end{align}
where $\underline{\mathbf{z}_{w}}\in \mathbb{R}^{N\times 1}$, $\underline{\boldsymbol{\epsilon}_{w}}\in \mathbb{R}^{N\times 1}$, and $\hat{\mathbf{F}}_{\mathrm{OOB}}\in\mathbb{C}^{K\times 1}$ are vectorized form of the finite-length signals $w[n]\cdot\underline{z^{(i)}}[n]$, $w[n]\cdot\underline{\epsilon^{(i)}}[n]$, and $\hat{F}_{\mathrm{OOB}}(e^{j\frac{2\pi k}{N}})$, respectively. The entry of matrix $\mathbf{V}\in\mathbb{C}^{K\times N}$ in the $k'$-th row and $n$-th column is $\frac{1}{\sqrt{N}}e^{-j\frac{2\pi n k}{N}}$, where $\frac{2\pi k}{N}$ is the $k'$-th discrete frequency in $(\rho\pi +\delta_{\mathrm{SL}},2\pi - \rho\pi - \delta_{\mathrm{SL}})$. Let $\underline{\mathbf{z}_{w,\mathcal{S}_i}}\in\mathbb{R}^{|\mathcal{S}_i|\times 1}$ and $\underline{\boldsymbol{\epsilon}_{w,\mathcal{S}_i}}\in\mathbb{R}^{|\mathcal{S}_i|\times 1}$ denote the subvectors of $\underline{\mathbf{z}_{w}}$ and $\underline{\boldsymbol{\epsilon}_{w}}$, respectively, whose indices $n$ are in $\mathcal{S}_{i}$. We can form the matrix $\mathbf{V}_{\mathcal{S}_i}\in\mathbb{C}^{K\times|\mathcal{S}_i|}$ by horizontally stacking the columns of $\mathbf{V}$ whose column indices are in $\mathcal{S}_i$. A numerical illustration of this construction is provided in Figure~\ref{fig:Vsi_illustration}. The pre-estimate of $\underline{\mathbf{z}_{w,\mathcal{S}_i}}$, denoted as $\underline{\tilde{\mathbf{z}}_{w,\mathcal{S}_i}}$, is then given by
\begin{align}\label{eq:windowed_modulo_preestimate}
    \underline{\tilde{\mathbf{z}}_{w,\mathcal{S}_i}} = \mathbf{V}_{\mathcal{S}_i}^{\dagger}\hat{\mathbf{F}}_{\mathrm{OOB}}.
\end{align}

\begin{figure*}[t!]
    \centering
    %\hspace*{-.5cm}
    \includegraphics[width = 1\textwidth]{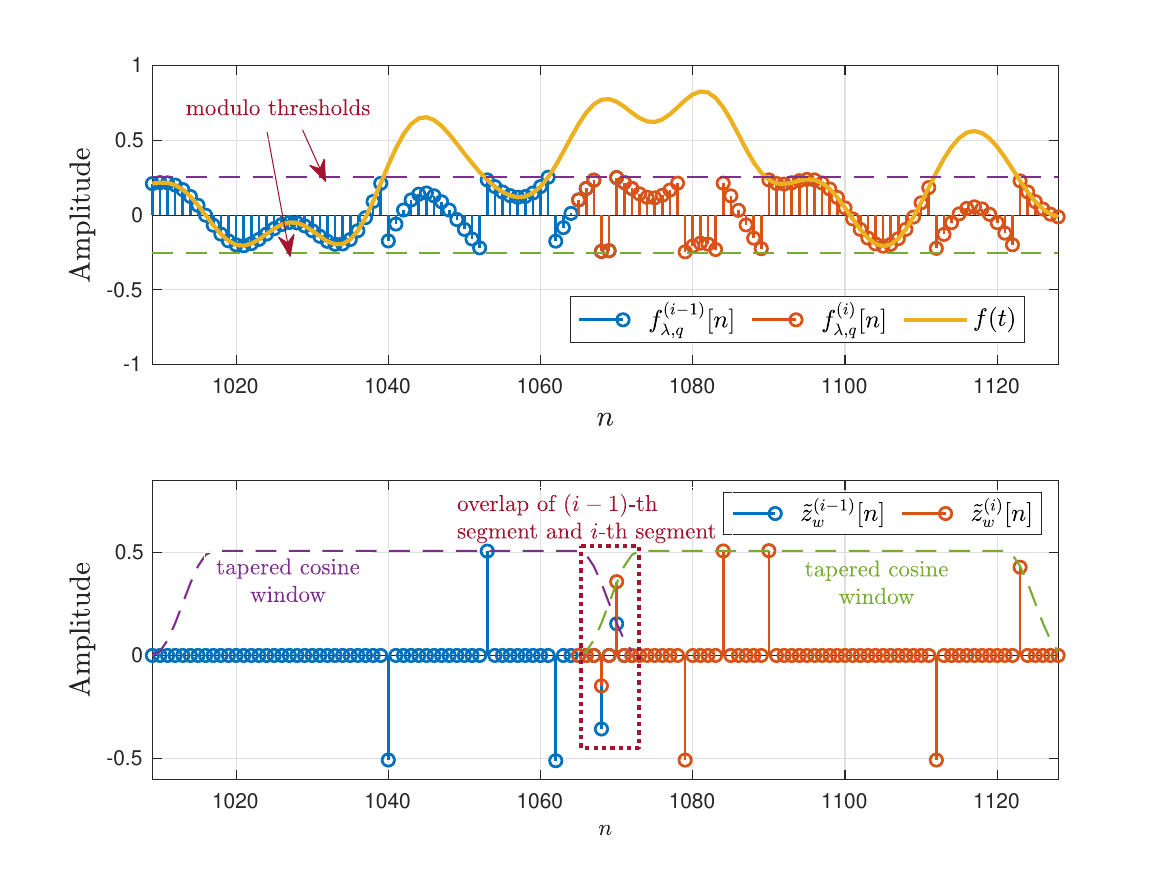}
    \caption{Illustration of the scaling correction for $\underline{\tilde{z}_{w}^{(i)}}[n]$. The top plot shows the original signal $f(t)$ and the folded samples of the modulo ADC for the $(i-1)$-th and $i$-th segments, denoted as $f_{\lambda',\mathrm{q}}^{(i-1)}[n]$ (blue) and $f_{\lambda',\mathrm{q}}^{(i)}[n]$ (red), respectively. The dashed constant lines correspond to the modulo thresholds. The bottom plot shows the first-order differences of the (windowed) modulo pre-estimates in the $(i-1)$-th and $i$-th segments, denoted $\underline{\tilde{z}_{w}^{(i-1)}}[n]$ (blue) and $\underline{\tilde{z}_{w}^{(i)}}[n]$ (red), respectively. Due to the tapered cosine window, the recovered amplitudes of $\underline{\tilde{z}_{w}^{(i-1)}}[n]$ and $\underline{\tilde{z}_{w}^{(i)}}[n]$ in the overlap region (dash-dot box) are attenuated. However, adding the last $\alpha N/2$ samples of $\underline{\tilde{z}_{w}^{(i-1)}}[n]$ to the first $\alpha N/2$ samples of $\underline{\tilde{z}_{w}^{(i)}}[n]$ corrects the amplitude of $\underline{\tilde{z}_{w}^{(i)}}[n]$.}
    \label{fig:recoveryDemo}
\end{figure*}

We obtain the (windowed) pre-estimate of $\underline{z^{(i)}}[n]$, denoted $\underline{\tilde{z}^{(i)}_{w}}[n]$, by placing the corresponding values of $\underline{\tilde{\mathbf{z}}_{w,\mathcal{S}_i}}$ to $\underline{\tilde{z}^{(i)}_{w}}[n]$ for $n\in\mathcal{S}_i$. The values of $\underline{\tilde{z}^{(i)}_{w}}[n]$ for $n\notin\mathcal{S}_i$ are set to 0. Note that the $n$-th element of $\underline{\tilde{z}^{(i)}_{w}}[n]$ for $n \in \left\{0,\cdots, \frac{\alpha N}{2}-1\right\}$ has a non-unity scaling of the desired modulo residue signal $\underline{z^{(i)}}[n]$ due to the tapered cosine window $w[n]$. To correct this scaling, we exploit the symmetry of the window function at the edges for $n\in\{0,1,\cdots,\frac{\alpha N}{2} - 1\}$:
\begin{align}
    w[n] + w[n + N\left(1-\frac{\alpha}{2}\right)] = 1.
    %w[n] = w\left[N\left(1-\frac{\alpha}{2}\right)-n\right].
\end{align}
We take the last $\frac{\alpha N}{2}$ samples of $\underline{\tilde{z}^{(i-1)}_{w}}[n]$ and add them to the first $\frac{\alpha N}{2}$ samples of $\underline{\tilde{z}^{(i)}_{w}}[n]$\footnote{Due to the STFT-based approach, $\tilde{z}^{(i-1)}_{w}[n]$ is already available when processing the $i$-th segment of $f_{\lambda',\mathrm{q}}[n]$}. The resulting signal is
\begin{align*}
   \underline{\tilde{z}^{(i)}}[n] = \begin{cases}
       \underline{\tilde{z}^{(i)}_{w}}[n] + \underline{\tilde{z}^{(i-1)}_{w}}\left[n + N\left(1 - \frac{\alpha}{2}\right)\right],\;\;0\leq n< \frac{\alpha N}{2}\\
        \qquad\qquad\;\;\underline{\tilde{z}^{(i)}_{w}}[n]\;\quad\quad\;,\;\;\frac{\alpha N}{2}\leq n < N\left(1-\frac{\alpha}{2}\right)
   \end{cases}
\end{align*}
The scaling correction is illustrated in Figure \ref{fig:recoveryDemo}. Since $z^{(i)}[n]\in 2\lambda'\mathbb{Z}$, $\underline{z^{(i)}}[n]\in 2\lambda'\mathbb{Z}$. The samples of the first-order difference of the modulo residue pre-estimates $\underline{\tilde{z}^{(i)}}[n]$ are rounded to the nearest integer multiple of $2\lambda'$ to obtain the first-order difference of the modulo residue estimates, denoted as $\underline{\hat{z}^{(i)}}[n]$. To undo the first-order difference operation in the modulo residue signal, the formula
\begin{align}\label{eq:remove_firstorder_diff}
    \hat{z}^{(i)}[n] = \hat{z}^{(i-1)}[n] + \underline{\hat{z}^{(i)}}[n]
\end{align}
is applied. The modulo residue signal estimate $\hat{z}^{(i)}[n]$ is then subtracted from $f_{\lambda',\mathrm{q}}^{(i)}[n]$ to unfold the $i$-th segment for $n\in\{0,\cdots, N(1-\frac{\alpha}{2})\}$. 

The recovery of the unfolded sequence continues until all the segments of $f_{\lambda',\mathrm{q}}[n]$ have been processed. If $c[n] = 0$ for all $n\in\{0,\cdots,N-1\}$ in the $i$-th segment, we can skip the processing of $i$-th segment and proceed to the next segment. Finally, a digital lowpass filter (LPF) with a passband region $\left(-\frac{\Omega_m}{2} - \delta_{\mathrm{SL}},+\frac{\Omega_m}{2} + \delta_{\mathrm{SL}}\right)$ is applied to the result, where $\Omega_m = \omega_m T_{\mathrm{s}}$. The unfolded signal per segment can be expressed as
\begingroup
\allowdisplaybreaks
\begin{align}\label{eq:unfolded_sig}
    \hat{f}^{(i)}[n] =& \left\{f_{\lambda',\mathrm{q}}^{(i)}[n] - \hat{z}^{(i)}[n]\right\}_{\mathrm{LPF}}\nonumber\\
    =&f^{(i)}[n] + \left\{z^{(i)}[n] - \hat{z}^{(i)}[n]\right\}_{\mathrm{LPF}} + \epsilon^{(i)}_{\mathrm{LPF}}[n],
\end{align}
\endgroup
where the notation $\{x[n]\}_{\mathrm{LPF}}$ indicates that the signal $x[n]$ is fed to the digital LPF.

\subsection{Design Considerations for Window Function}
\label{subsection:window_design}
\begin{figure*}[!t]
    \centering
    \subfloat[ ]{
        \includegraphics[width=0.485\linewidth]{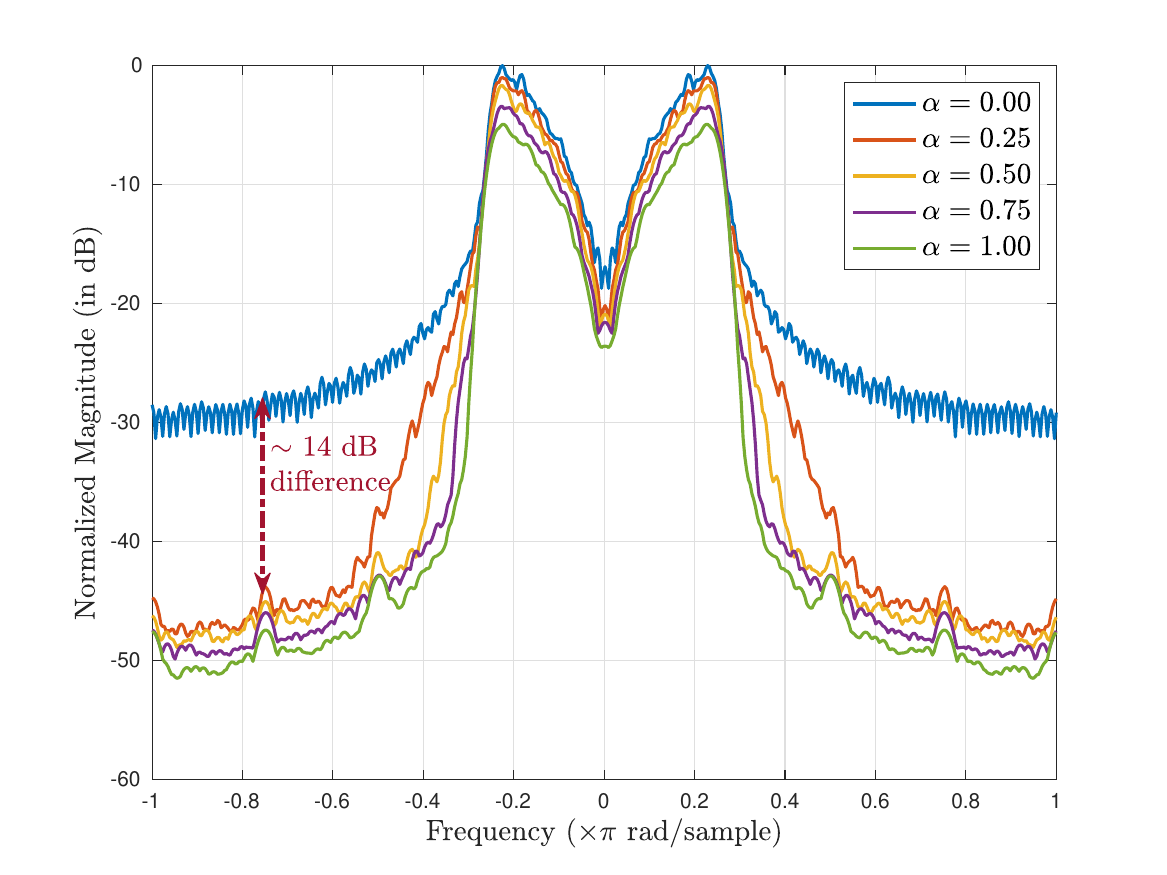}
        \label{fig:spectral_leakage_N64}
    }
    \hfill
    \subfloat[ ]{
        \includegraphics[width=0.485\linewidth]{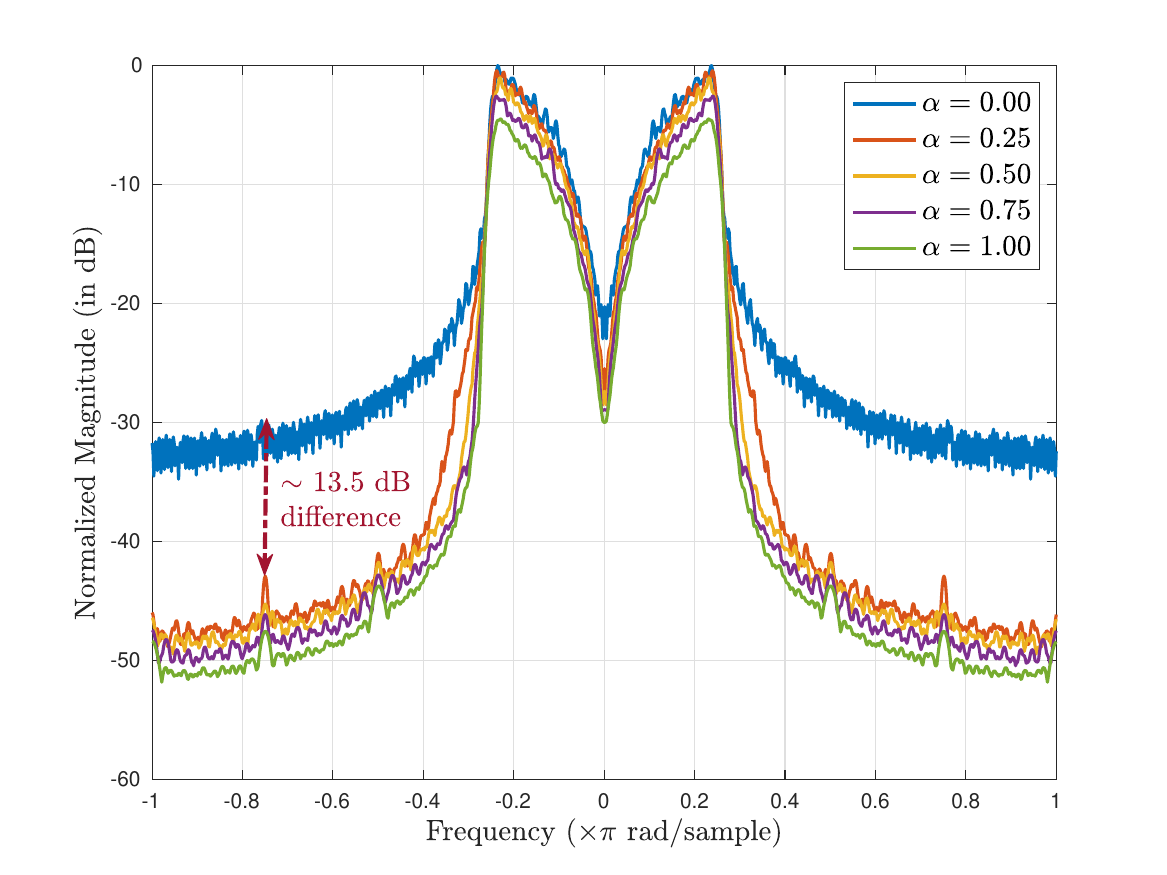}
        \label{fig:spectral_leakage_N128}
    }
    \caption{Normalized magnitude (in dB) vs. angular frequency of $w[n]\cdot \underline{f}[n]$ for different values of $\alpha$ with (a) $N = 64$ and (b) $N = 128$. The original signal $f(t) = \sum_{i = 0}^{150,000}\mathrm{sinc}(t - i)$ is sampled at four times the Nyquist rate to get $f[n]$. The magnitude response is computed using a max-hold across the zero-padded DFTs of all length-$N$ frames.}
    \label{fig:overall}
\end{figure*}

The design of the window function is crucial for the proposed STFT-based recovery algorithm because the unfolding procedure depends on the effective isolation of the out-of-band modulo residue from the in-band signal components. A rectangular window provides the narrowest mainlobe and thus offers the best frequency resolution, but its large sidelobes cause substantial spectral leakage. This leakage is problematic in our setting because it can mask or distort the relatively small out-of-band residue terms used for unfolding. To balance resolution and leakage suppression, we adopt a tapered cosine window. The tapered cosine window preserves a relatively high frequency resolution while offering much lower sidelobes. Moreover, its roll-off parameter allows us to control the trade-off between mainlobe width and sidelobe attenuation. This makes it well suited for separating the in-band and out-of-band components necessary for accurate recovery.

To illustrate this trade-off, we consider the windowing of an oversampled train of sinc pulses. The impact of the roll-off factor $\alpha$ on the spectral leakage of $w[n]\cdot \underline{f}[n]$ for $N = 64$ and $N = 128$ is shown in Figures \ref{fig:spectral_leakage_N64} and \ref{fig:spectral_leakage_N128}, respectively. As shown in the plots, there is a pronounced difference between the out-of-band energy produced by a rectangular window ($\alpha = 0$) and that of a tapered cosine window with $\alpha = 0.25$. Increasing $\alpha$ further results in a faster sidelobe decay and improved suppression of out-of-band components.

The choice of the window length $N$ and roll-off parameter $\alpha$ also introduces important design trade-offs. Increasing $N$ improves the frequency resolution but lengthens the STFT frame, thereby increasing the computational complexity of the algorithm. Meanwhile, the roll-off factor $\alpha$ governs how fast the window tapers: larger $\alpha$ yields better sidelobe suppression and reduced leakage, but at the cost of a wider mainlobe and lower frequency resolution; smaller $\alpha$ behaves more like a rectangular window, improving resolution but reintroducing leakage. Furthermore, $\alpha$ dictates the amount of overlap between consecutive segments and the hop size of the algorithm. Thus, the proposed algorithm tends to have a slightly higher computational complexity for a larger $\alpha$ because more overlapping segments must be processed. The impact of $N$ and $\alpha$ on the computational complexity of the proposed algorithm is established in Section \ref{subsection:complexity}. In practice, these window parameters must be chosen jointly to achieve sufficient suppression of spectral leakage while maintaining a manageable computational overhead.

%keeping computational overhead manageable.

\subsection{Proposed Recovery Scheme vs. Other Fourier-based Reconstruction Techniques}

Because of the STFT computation, our proposed recovery scheme can be classified as a Fourier-based recovery technique. Other algorithms under this category are the Fourier-Prony method \cite{Fourier-Prony}, USLSE \cite{Zhang:2023}, $B^2R^2$ \cite{Azar:2022}, and LASSO-$B^2R^2$ \cite{Shah:2023}. The advantage of our proposed recovery scheme is that it does not restrict the signal to decay over time (such as in $B^2R^2$ and LASSO-$B^2R^2$) nor does it require the signal to be periodic (such as in the Fourier-Prony method). The window function ensures that sharp discontinuities are avoided at the periodic boundaries when the length-$N$ sampled signal is repeated. Consequently, spectral leakage is minimized. The vanishing signal amplitude requirement in $B^2R^2$ and LASSO-$B^2R^2$ may require a long observation time to estimate modulo residue samples. The computational complexities of these algorithms increase rapidly with the observation length. In contrast, the proposed algorithm follows the principle of STFT and the observation length of a segment can be controlled. As such, we can either choose to unfold few overlapping segments with large observation length or many overlapping segments with small observation length. In addition, the STFT-based approach enables processing of modulo samples in short frames. This is suitable for real-time applications with low-latency requirements.

Our objective is to derive the MSE performance guarantees and computational complexity for this recovery scheme and demonstrate the advantage of modulo ADCs that use the proposed recovery scheme over conventional ADCs. The main results of our paper are presented in the next section.

%In the next section, the main results of our paper are presented.

\section{Performance Guarantees using the Proposed Recovery Procedure}  
\label{section:performance_guarantee}

In this section, we establish theoretical performance guarantees for the proposed STFT-based signal recovery method. We begin by introducing the primary performance metric, the mean squared error (MSE), and justify the statistical model for the dither signal and resulting quantization noise. Next, we define the parameters that capture the effects of (non-negligible) spectral leakage in the DFT computation. Building on these foundations, we present sufficient conditions for the oversampling factor and quantization resolution to ensure accurate unfolding of modulo residues. We also compare \textcolor{black}{the performance of the proposed STFT-based recovery method} with that of conventional ADCs and existing modulo ADC recovery techniques. The section concludes with a discussion of the proposed algorithm's computational complexity.

\subsection{Measuring Recovery Performance}
Our primary measure of the recovery performance of the unfolding scheme is the MSE between the (sampled) input signals $f[n]$ and the $\hat{f}[n]$, given by

\begingroup
\allowdisplaybreaks
\begin{align}\label{eq:MSE}
    \mathrm{MSE} =& \frac{1}{N_0}\sum_{n = 0}^{N_0-1}\mathbb{E}\left\{\big|\hat{f}[n] - f[n]\big|^2\right\}\nonumber\\
    =& \frac{1}{NI}\sum_{i=1}^{I}\sum_{n = 0}^{N-1}\mathbb{E}\left\{\big|\hat{f}^{(i)}[n] - f^{(i)}[n]\big|^2\right\}\nonumber\\
    =& \frac{1}{NI}\sum_{i=1}^{I}\sum_{n = 0}^{N-1}\bigg|\left\{z^{(i)}[n] - \hat{z}^{(i)}[n]\right\}_{\mathrm{LPF}}\bigg|^2\nonumber\\
    &+2\mathbb{E}\left\{\epsilon^{(i)}_{\mathrm{LPF}}[n]\right\}\left\{z^{(i)}[n] - \hat{z}^{(i)}[n]\right\}_{\mathrm{LPF}}\nonumber\\
    &+\mathbb{E}\left\{\big|\epsilon^{(i)}_{\mathrm{LPF}}[n]\big|^2\right\}\nonumber\\
    =& \frac{1}{NI}\sum_{i=1}^{I}\sum_{n = 0}^{N-1}\bigg|\left\{z^{(i)}[n] - \hat{z}^{(i)}[n]\right\}_{\mathrm{LPF}}\bigg|^2\nonumber\\
    &\;+\mathbb{E}\left\{\big|\epsilon^{(i)}_{\mathrm{LPF}}[n]\big|^2\right\},
\end{align}
\endgroup
where $N_0$ denotes the signal length. The second line was obtained from partitioning the original signal into $I$ length-$N$ segments. The third line follows from the expression of the unfolded signal in equation \eqref{eq:unfolded_sig} and from performing some algebraic manipulation. The fourth line comes from the fact that the quantization noise has zero mean. Thus, the second term on the third line can be removed. 

From equation \eqref{eq:MSE}, it can be seen that the MSE expression comes from two sources: (1) the error due to the modulo residue estimation error and (2) the in-band quantization noise. Our objective in this section is to derive the exact MSE expression as a function of $\mathrm{OF}$ and $b$ when the proposed recovery procedure is used.

\subsection{Statistical Model for Quantization Noise}\label{subsection:stat_quant_noise}

Since MSE involves the mean square of the quantization noise, it is crucial to understand the statistical properties of the quantization noise. The statistical nature of the quantization noise originates from the dithered quantization framework established in \cite{Gray:1993}. This framework is appropriate for modulo ADC because the folding operation ensures that the signal does not overload the quantizer. The `no overloading' property of the modulo ADC, together with the dither signal $d[n]$, guarantees that the quantization noise is a white process.

In our problem setup, we use a triangular dither rather than a uniform dither because of its favorable second-order statistical properties. In particular, the power of the quantization noise induced by the triangular dither is independent of the input, i.e.,
\begin{align}
    \mathbb{E}\left\{|\epsilon[n]|^2\big|f_{\lambda'}[n]\right\} = \mathbb{E}\left\{|\epsilon[n]|^2\right\}
\end{align}
as established in \cite[Theorem 2]{Gray:1993}. This input-independence of the second-order statistics of quantization noise does not generally hold for a single uniform dither \cite{Gray:1993}. For the triangular dither $d[n]$ described in Section~\ref{section:sys_model}, the resulting quantization noise power is
\begin{align}
    \mathbb{E}\left\{|\epsilon[n]|^2\right\} = \frac{1}{4}\left(\frac{2\lambda}{2^b}\right)^2 = \frac{\lambda^2}{2^{2b}}
\end{align}

\subsection{Incorporating Spectral Leakage}

Since spectral leakage cannot be completely neglected for small \( N \), we explicitly accounted for the effect of non-negligible spectral leakage in the proposed recovery algorithm. Thus, we define the \emph{spectral leakage bin count} as
\begin{align}
    K_{\mathrm{SL}} = 2\bigg\lceil\frac{\delta_{\mathrm{SL}} \cdot N}{\pi}\bigg\rceil,
\end{align}
which corresponds to the number of discrete frequency bins within the spectral leakage interval \( (\rho\pi, \rho\pi + \delta_{\mathrm{SL}}) \cup (2\pi - \rho\pi - \delta_{\mathrm{SL}}, 2\pi - \rho\pi) \). The frequency-domain plots in Figure \ref{fig:Vsi_illustration} provides a visual illustration of the intuition behind the spectral leakage bin count. Although \( K_{\mathrm{SL}} \) and \( \delta_{\mathrm{SL}} \) are generally difficult to compute exactly in practice due to their dependence on the precise shape of the leakage spectrum, we highlight their role in the MSE performance guarantee to reveal how spectral leakage affects recovery accuracy. Finally, we note that the MSE performance guarantee to be derived does not account for the spectral leakage inside $[\rho\pi+\delta_{\mathrm{SL}},2\pi-\rho\pi - \delta_{\mathrm{SL}}]$. Nonetheless, our numerical results in Section \ref{section:numerical_results} demonstrate that our theoretical MSE predictions still coincide with the simulated MSE despite neglecting some spectral leakage.

\subsection{Main Result}
Before we state the main result of this paper, a proposition about the matrix $\mathbf{V}_{\mathcal{S}_i}$ is presented. This plays a key role in deriving the main result.

\begin{proposition}\label{proposition:oversampling_req_gen}
    Let $f(t)$ be a bandlimited function. The matrices $\{\mathbf{V}_{\mathcal{S}_i}\}_{i}$ used in the recovery algorithm have full column rank if 
\begin{align}\label{eq:OF_requirement}
        \mathrm{OF}\geq 
        \frac{N}{N - \underset{i}{\max}\;|\mathcal{S}_i| - K_{\mathrm{SL}}}.
    \end{align} 
\end{proposition}
\begin{proof}
    See Appendix \ref{proof:appendix_A}.
\end{proof}
With $\mathbf{V}_{\mathcal{S}_{i}}$ being full column rank, 
$\mathbf{V}_{\mathcal{S}_i}^\dagger\mathbf{V}_{\mathcal{S}_i}$ is a $|\mathcal{S}_i|\times|\mathcal{S}_i|$ identity matrix. Under a `no quantization noise setting', $\|\underline{\tilde{\mathbf{z}}_{\mathcal{S}_i}} - \underline{\mathbf{z}_{\mathcal{S}_i}}\|_{\infty} = \|\mathbf{V}_{\mathcal{S}_i}^\dagger\mathbf{V}_{\mathcal{S}_i}\underline{{\mathbf{z}}_{\mathcal{S}_i}} - \underline{\mathbf{z}_{\mathcal{S}_i}}\|_{\infty} = 0$. Thus, the full column rank property is crucial for the recovery of $z^{(i)}[n]$.

One key observation in Proposition \ref{proposition:oversampling_req_gen} is that the oversampling factor $\mathrm{OF}$ approaches unity if $|\mathcal{S}_{i}|$ grows strictly slower with $N$. This is true for finite-energy bandlimited functions due to their time-domain decay property (see \cite{Shah:2023,Azar:2022}). That is, $\exists n_0$ such that $f[n] < \lambda'$ for all $|n| > n_0$. Consequently, $\exists N_0$ such that $|\mathcal{S}_i|$ does not change $\forall N > N_0$. Thus, the sampling rate can be made closer to the Nyquist rate for finite-energy bandlimited signals at the expense of a longer observation window. Similar trend was observed in the recovery guarantees established in \cite{Beckmann:2024}. That is, for signals that are compact $\lambda$-exceedance with parameter $\rho_{e}$, smaller $\mathrm{OF}$ can be used by increasing the number of observations. This $\lambda$-exceedance property is closely related to the time-domain decay property in \cite{Shah:2023,Azar:2022}.

One drawback of Proposition \ref{proposition:oversampling_req_gen} is that it depends on the bandlimited input signal $f(t)$. The following proposition provides a sufficient condition for $\mathrm{OF}$ that depends only on the maximum amplitude $\|f(t)\|_{\infty}$ and not on the entire signal $f(t)$.
\begin{proposition}\label{proposition:oversampling_req}
    The matrices $\{\mathbf{V}_{\mathcal{S}_i}\}_{i}$ used in the recovery algorithm have full column rank if 
\begin{align}\label{eq:OF_requirement}
        \mathrm{OF}\geq \frac{3}{1- \frac{K_{\mathrm{SL}}}{N}}.
    \end{align} 
This is achieved by setting the modulo threshold to 
\begin{align}
    \lambda' = \frac{\|f(t)\|_{\infty}}{\mathrm{OF}\cdot\left[1-\frac{K_{\mathrm{SL}}}{N}\right] - 2}.
\end{align}
%\[\lambda' = \min\left(\|f(t)\|_{\infty},\frac{\|f(t)\|_{\infty}}{\mathrm{OF}
%\left(1-\frac{K_{\epsilon}(\alpha,N)}{N}\right) - 2}\right).\]
\end{proposition}
\begin{proof}
    See Appendix \ref{proof:appendix_B}.
\end{proof}
Note that Proposition \ref{proposition:oversampling_req} is sufficient, but not a necessary condition. The expression used to bound $\max_{i}|\mathcal{S}_i|$ from above is identical to the bound of the folding instances derived in \cite[Equation 8]{Shah:2023}. However, their proof assumed that $f(t)$ is a bandlimited signal with negligible truncation error. We show in Appendix \ref{proof:appendix_B} that this bound on $\max_{i}|\mathcal{S}_i|$ is also valid for the setup considered in this paper. However, this bound can be loose in some cases as demonstrated in \cite[Table 1]{Shah:2023}.

For $N$ approaching infinity, equation \eqref{eq:OF_requirement} simplifies to $\mathrm{OF} > 3$, which was derived in \cite{Bernardo_ISIT2024}. The theoretical MSE guarantee established in \cite{Bernardo_ISIT2024} does not take into account the spectral leakage. As demonstrated in Proposition \ref{proposition:oversampling_req}, the spectral leakage bin count $K_{\mathrm{SL}}$ increases the $\mathrm{OF}$ required to satisfy condition \eqref{eq:OF_requirement}. Nonetheless, its effect on the OF sufficient condition diminishes fast as $N$ grows. 

We now state the main theoretical result of this paper. The following theorem establishes the MSE of the proposed recovery method when parameters $\mathrm{OF}$ and $b$ are above certain values.
\begin{theorem}\label{theorem:MSE_guarantee}
   Suppose 
   %\textcolor{black}{the constant offset $2\lambda'm$ is known,}
    \begin{align}
        \mathrm{OF} \geq \frac{3}{1-\frac{K_{\mathrm{SL}}}{N}}
    \end{align}
    and
    \begin{align}\label{eq:b_req}
        b > 3 + \log_2\left(1+\frac{3M}{4}\right),
    \end{align}
    %\[\mathrm{OF} \geq \frac{3}{1-\frac{K_{\mathrm{SL}}}{N}}\;\text{ and }\;b > 3 + \log_2\left(1+\frac{3M}{4}\right),\]
    where $M = \underset{i}{\max}\;\|\mathbf{V}_{S_i}^\dagger\mathbf{V}_{S_i^c}\|_{\infty}$. The MSE incurred by the proposed recovery algorithm can be written as
    \begin{align}\label{eq:MSE_guarantee}
        \mathrm{MSE} = \frac{\|f(t)\|_{\infty}^2 \left(1+\frac{\delta_{\mathrm{SL}}}{\pi}\cdot\mathrm{OF}\right)}{\mathrm{OF}\cdot(2^b-2)^2\cdot\left(\mathrm{OF}\left[1-\frac{K_{\mathrm{SL}}}{N}\right]-2\right)^2}.
    \end{align}
    This is achieved by setting the modulo threshold to 
\begin{align}\label{eq:lambda_p_req}
    \lambda' = \frac{\|f(t)\|_{\infty}}{\mathrm{OF}\cdot\left[1-\frac{K_{\mathrm{SL}}}{N}\right] - 2}.
\end{align}
\end{theorem}
\begin{proof}
    See Appendix \ref{proof:appendix_C}.
\end{proof}
The intuition behind this result is that the sufficient conditions for $\mathrm{OF}$ and $b$ in Theorem \ref{theorem:MSE_guarantee} guarantee that the bounded noise added to the modulo residue pre-estimate $\tilde{z}^{(i)}[n]$ is within $(-\lambda',+\lambda')$. Since $\tilde{z}^{(i)}[n]$ are integer multiples of $2\lambda'$, the rounding operation maps all modulo residue pre-estimates $\tilde{z}^{(i)}[n]$ onto the correct modulo residue $z^{(i)}[n]$. Consequently, only the quantization noise power term in equation \eqref{eq:MSE} appears in the MSE. 

The derivation of  \cite[Theorem 1]{Bernardo_ISIT2024} did not consider the contribution of the $\epsilon^{(i)}[n]$ for $n\in\mathcal{S}_i^c$, resulting in $b > 3$ condition. In Theorem \ref{theorem:MSE_guarantee}, the second term in equation \eqref{eq:b_req} accounts for the impact of $\epsilon_{\mathcal{S}_i^c}^{(i)}[n]$ on recovery performance. Although we are interested in the modulo residue samples located at $\mathcal{S}_{i}$, $\epsilon_{\mathcal{S}_i^c}^{(i)}[n]$ still contributes to $\hat{z}^{(i)}[n]$ through the computation of $\hat{F}_{\mathrm{OOB}}(e^{j\frac{2\pi k}{N}})$. In Section \ref{section:numerical_results}, we investigate how the second term of \eqref{eq:b_req} is affected by parameters such as segment length, cardinality of $\mathcal{S}_i$, and oversampling factor.

It is also important to point out the detrimental impact of spectral leakage on MSE performance. Neglecting the spectral leakage, that is, $\delta_{\mathrm{SL}} = 0$, equation \eqref{eq:MSE_guarantee} simplifies to
\begin{align}
    \mathrm{MSE} &= \frac{\|f(t)\|_{\infty}^2 }{\mathrm{OF}\cdot(2^b-2)^2\cdot\left(\mathrm{OF}-2\right)^2}\nonumber\\
    &=\mathcal{O}\left(\frac{1}{\mathrm{OF}^3}\right)
\end{align}
This is similar to the asymptotic growth rate established in \cite{Bernardo_ISIT2024} which neglected spectral leakage. When $\delta_{\mathrm{SL}} > 0$, the asymptotic growth rate of MSE becomes $\mathcal{O}\left(\frac{1}{\mathrm{OF}^2}\right)$ only. $\delta_{\mathrm{SL}}$ can be reduced by using a larger window length. However, using a larger window length increases the computational complexity of sliding window DFT (see Section \ref{subsection:complexity}).

\subsection{Comparison with Recovery Guarantees of Existing Methods}\label{subsection:compare_recovery}

We now compare the derived theoretical results with other performance guarantees derived for modulo sampling. A popular recovery procedure is the higher-order differences (HoD) approach developed in \cite{Bhandari:2021}. Performance guarantee for this algorithm under a bounded noise setting (e.g., quantization noise) was analyzed in \cite[Theorem 6]{Bhandari:2021}. More precisely, they showed that for finite-energy bandlimited signals, noisy unfolded samples can be recovered from noisy modulo samples up to an unknown additive constant, that is,
\begin{align}
    \hat{f}_{\mathrm{HoD}}[n] = f[n] +\epsilon[n]+2\lambda'p,
\end{align}
where $p\in\mathbb{Z}$ is unknown, if the sampling rate is at least $2^{\alpha}\pi e\times f_{\mathrm{Nyq}}$. Here, $\alpha\in\mathbb{N}$ is a parameter of the bounded noise and depends on both the maximum amplitude of the noise and modulo threshold. In contrast, Theorem \ref{theorem:MSE_guarantee} only requires $\mathrm{OF} > 3$ under negligible spectral leakage. We also note that $\mathrm{OF} > 3$ closely resembles the OF requirement derived in \cite{Zhang:2024} to uniquely identify periodic bandlimited signals under the modulo-DFT sensing model. However, their model first applies DFT to the sampled signal prior to modulo operation. Moreover, the $\mathrm{OF} > 3$ result is equivalent to Itoh's condition \cite{Itoh:1982} and is reminiscent of the spatial oversampling requirement established for DoA estimation using MIMO arrays with modulo nonlinearities \cite{Fernandez-Menduina:2022}.

A prediction-based approach \cite{Romanov:2019} showed that $\mathrm{OF}$ can be made arbitrarily close to unity in exchange for a significant increase in the prediction filter length. However, it was pointed out in \cite[Section III]{Romanov:2019} that their proposed approach \emph{``collapses in the presence of quantization noise"}. In contrast, our algorithm can operate with finite quantization bits. Moreover, our proposed algorithm \textcolor{black}{works at sampling rates} arbitrarily close to the Nyquist rate for finite-energy bandlimited signals and a sufficiently large window length $N$ by virtue of Proposition \ref{proposition:oversampling_req_gen}. We also note that the $\mathrm{OF}$ requirement established in \cite{Fourier-Prony} for periodic bandlimited signals is
\begin{align}
    \mathrm{OF} \geq \frac{N}{N - 2|\mathcal{S}| - 2},
\end{align}
where $|\mathcal{S}|$ denotes the number of folding instances. Evidently, when $K_{\mathrm{SL}} = 0$, the sufficient condition for the $\mathrm{OF}$ established in Proposition 1 is better than the $\mathrm{OF}$ requirement in \cite{Fourier-Prony}. The advantage of Proposition 1 might be attributed to the availability of 1-bit folding information in our setup.

Finally, there exists a line of work \cite{Florescu:2022,Florescu:2022b,Geethu:2025} that uses time-domain thresholding to unfold the output produced by generalized modulo sampling models. Unlike DFT-based recovery methods, thresholding is performed in the time domain; therefore, spectral leakage is not an issue. Moreover, thresholding-based techniques have established recovery guarantees \cite{Florescu:2022,Florescu:2022b}. However, the sequential estimation of the folding parameters requires that the folding times be well-separated in the time domain. Such a separation cannot be guaranteed under ideal modulo sampling with zero hysteresis as demonstrated in \cite[Lemma 1]{Florescu:2022b}. Moreover, the MSE bound established in \cite[Proposition 1]{Florescu:2022b} is inversely proportional to the hysteresis $h$. Under ideal modulo nonlinearities, this $h$ goes to zero, and thus the MSE under the thresholding-based recovery becomes unbounded.

\subsection{Comparison with Conventional ADCs}

To demonstrate the advantage of a modulo ADC with 1-bit folding information over a conventional ADC, we first derive the MSE of a conventional $b$-bit ADC under a non-subtractive dithered quantization framework. We compared the modulo ADC and the conventional ADC under the same bit budget used for amplitude quantization. For a triangle dither $d[n] \in\left(-\frac{2\lambda}{2^b},+\frac{2\lambda}{2^b}\right)$, the ADC parameter $\lambda$ should be set to $\lambda = \left(1+\frac{1}{2^b}\right)\|f(t)\|_{\infty}$. After digital filtering, the quantization noise power (which is also the MSE) becomes
\begin{align}\label{eq:MSE_conv}
    \mathrm{MSE}_{\mathrm{conv}} = \frac{\|f(t)\|_{\infty}^2}{\mathrm{OF}(2^b-2)^2}.
\end{align}

Comparing the MSE guarantee for modulo ADC in Theorem \ref{theorem:MSE_guarantee} and the derived MSE for a conventional ADC, it can be observed that $\mathrm{MSE}_{\mathrm{mod}} = \mathcal{O}\left(\frac{1}{\mathrm{OF}^2}\right)$ while $\mathrm{MSE}_{\mathrm{conv}} = \mathcal{O}\left(\frac{1}{\mathrm{OF}}\right)$ under sufficiently large $b$. The fast decay rate of $\mathrm{MSE}_{\mathrm{mod}}$ with respect to $\mathrm{OF}$ is due to the reduction of the ADC parameter $\lambda$. A smaller ADC range results in smaller quantization bins. This demonstrates the superior performance of modulo ADCs compared to conventional ADCs in oversampled systems. However, the modulo ADC has a slightly higher hardware complexity due to the folding operation and the 1-bit folding information $c[n]$. A detailed comparison of power consumption, cost, and temperature behavior requires specific circuit-level implementations of modulo ADCs and conventional ADCs, which is beyond the scope of this simulation-based study.

\subsection{Computational Complexity}
\label{subsection:complexity}
We now derive the computational complexity of the proposed algorithm in terms of signal length $N_0$, window length $N$, window roll-off $\alpha$ ($0\leq \alpha \leq 1$) and oversampling factor $\mathrm{OF} = \frac{1}{\rho}$. The number of segments depends on the ratio $\frac{N_0}{N}$, and parameter $\alpha$ dictates the extent of overlap between consecutive segments. For a large $N_0$, the number of segments to be processed by the algorithm is in $\mathcal{O}\left(\frac{N_0}{N\left(1-\alpha/2\right)}\right)$. Considering the complexity of the residual recovery algorithm in each frame, the calculation of $\mathbf{V}_{\mathcal{S}_i}^\dagger \in \mathbb{C}^{|\mathcal{S}_i|\times K}$ is the most computationally-expensive operation. More precisely, this calculation is in $\mathcal{O}\left(K|\mathcal{S}_{\max}|^2\right)$, where $|\mathcal{S}_{\max}| = \max_{i}|\mathcal{S}_i|$. Since $K \approx (1-\rho)N$ and $|\mathcal{S}_{\max}| \leq K$ under appropriate choice of $\mathrm{OF}$ and $\lambda'$, the complexity of computing $\mathbf{V}_{\mathcal{S}_i}^\dagger$ is in $\mathcal{O}\left((1-\rho)^3N^3\right)$. Considering all segments to be processed, the overall computational complexity of the proposed recovery method is in $\mathcal{O}\left(\frac{N_0 (1-\rho)^3N^2}{1-\alpha/2}\right)$. Hence, the computational complexity of the proposed algorithm is linear in the signal length $N_0$.

Based on the derived computational complexity of the proposed algorithm, the speed of the proposed STFT-based recovery scheme can be improved by selecting a small oversampling factor (high $\rho$) that satisfies the sufficient condition in Theorem \ref{theorem:MSE_guarantee}. However, reducing $\mathrm{OF}$ negatively impacts the MSE as shown in equation \eqref{eq:MSE_guarantee}. The speed of the proposed STFT-based recovery scheme can also be improved by selecting a small window length $N$. However, a small window length also requires a higher $\alpha$ to compensate for the increased spectral leakage. Finally, we note that the recovery algorithm in \cite{Bernardo_ISIT2024,Bernardo2025TSP} is a special case of our proposed recovery algorithm with $N = N_0$ and $\alpha = 0$, that is, a rectangular window with no overlap. Hence, its complexity is in $\mathcal{O}((1-\rho)^3N_0^3)$. Because $N$ is typically chosen to be much smaller than $N_0$, our proposed STFT-based recovery is $\mathcal{O}\left(\frac{N_0^2(1-\alpha/2)}{N^2}\right)$ faster than the recovery algorithm in \cite{Bernardo_ISIT2024}. Suppose that the signal to be processed has $N_0 = 100,000$ samples. Our proposed algorithm with a sliding window length $N = 256$ and roll-off parameter $\alpha = 0.25$ can unfold the whole signal at roughly $7.48\times10^{-6}$ times the total time needed by the recovery method in \cite{Bernardo_ISIT2024,Bernardo2025TSP} to unfold the entire signal.

\section{Numerical Results}
\label{section:numerical_results}

In this section, we validate the MSE performance guarantees for the proposed algorithm established in the previous section by comparing the theoretical MSE results with the simulated performance. Numerical experiments are also conducted to gain additional insight on how the algorithm parameters affect the performance guarantees. We also compare the proposed STFT-based recovery method with the HoD-based recovery. 

\subsection{Analysis of the Sufficient Condition for ADC Resolution}

We first investigate how the sufficient condition on $b$ (i.e., equation \eqref{eq:b_req}) is affected by the algorithm and input signal parameters such as segment length $N$, oversampling factor $\mathrm{OF}$, and number of non-zero elements of the 1-bit folding information signal $c[n]$. To this end, we consider three $\mathrm{OF}$ parameter settings ($\mathrm{OF} = 4,\;8,\;12$) and three segment lengths ($N = 64,\;128,\;256$). We also consider three values of $|\mathcal{S}_i|$ ($|\mathcal{S}_i| = \frac{N}{32},\frac{N}{16},\frac{N}{8}$). The selected window lengths and oversampling factor settings are chosen arbitrarily and are intended solely to demonstrate how these parameters affect the second term of the LHS of equation \eqref{eq:b_req}. For each combination of these parameter settings, we generate 100,000 realizations of $\mathbf{V}_{\mathcal{S}_i}$ by selecting $|\mathcal{S}_i|$ out of $N$ columns of $\mathbf{V}$. We then estimate the second term of equation \eqref{eq:b_req} by calculating
\begin{align}
    \tilde{M} = \max_{i\in\{1,\;2,\cdots,\;100,000\}} \|\mathbf{V}^\dagger_{\mathcal{S}_i}\mathbf{V}_{\mathcal{S}_i^c}\|_\infty,
\end{align}
where the index $i$ iterates over all 100,000 random realizations of $\mathbf{V}_{\mathcal{S}_i}$.

\begin{figure}[t!]
    \hspace*{-.25cm}
    \includegraphics[scale = .48]{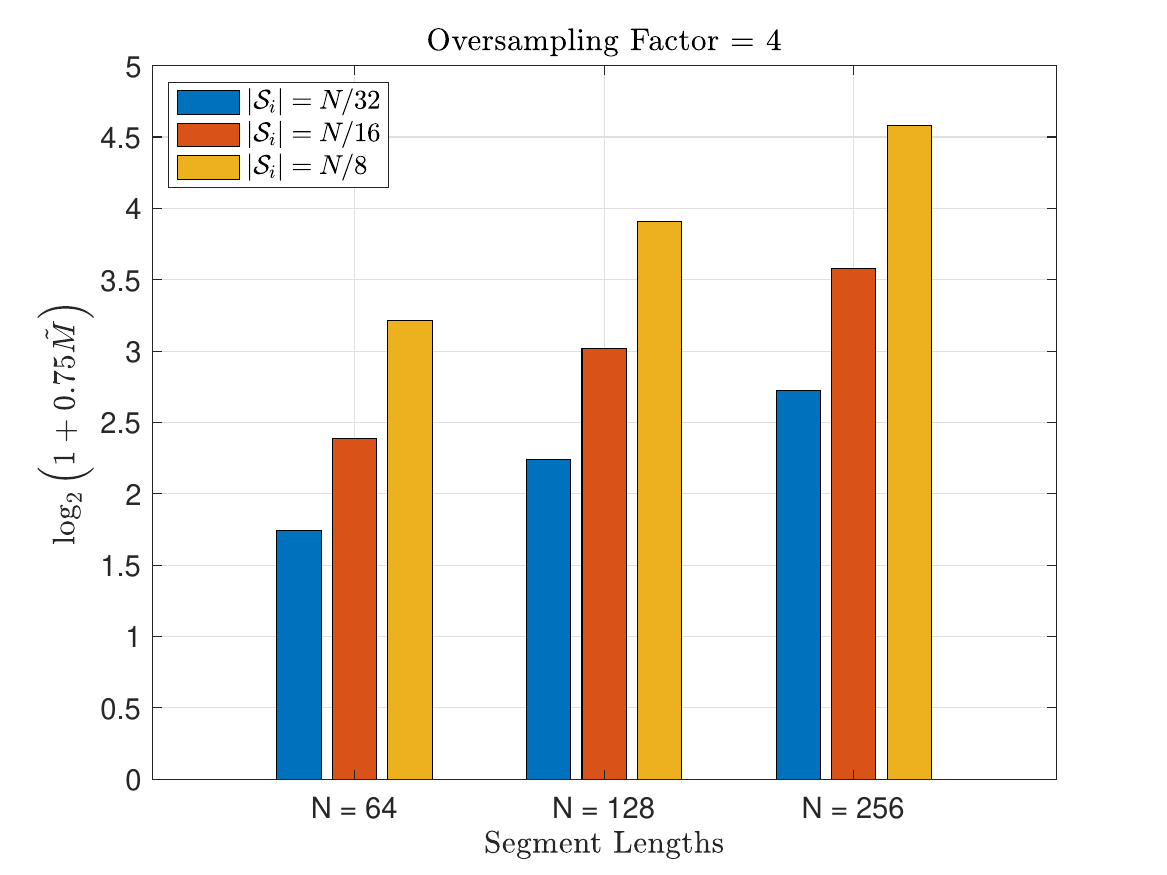}
    \caption{Empirical evaluation of $\log_2\left(1+0.75M\right)$ as a function of segment length ($N$) and $|\mathcal{S}_i|$ for $\mathrm{OF} = 4$. \textcolor{black}{The values of the segment length used in the numerical experiments are $N = 64$, $128$, and $256$. The values of $|\mathcal{S}_i|/N$ are set to $\frac{1}{8}$, $\frac{1}{16}$, and $\frac{1}{32}$.}}
    \label{fig:V_norm_OF4}
\end{figure}
\begin{figure}[t!]
    \hspace*{-.25cm}
    \includegraphics[scale = .49]{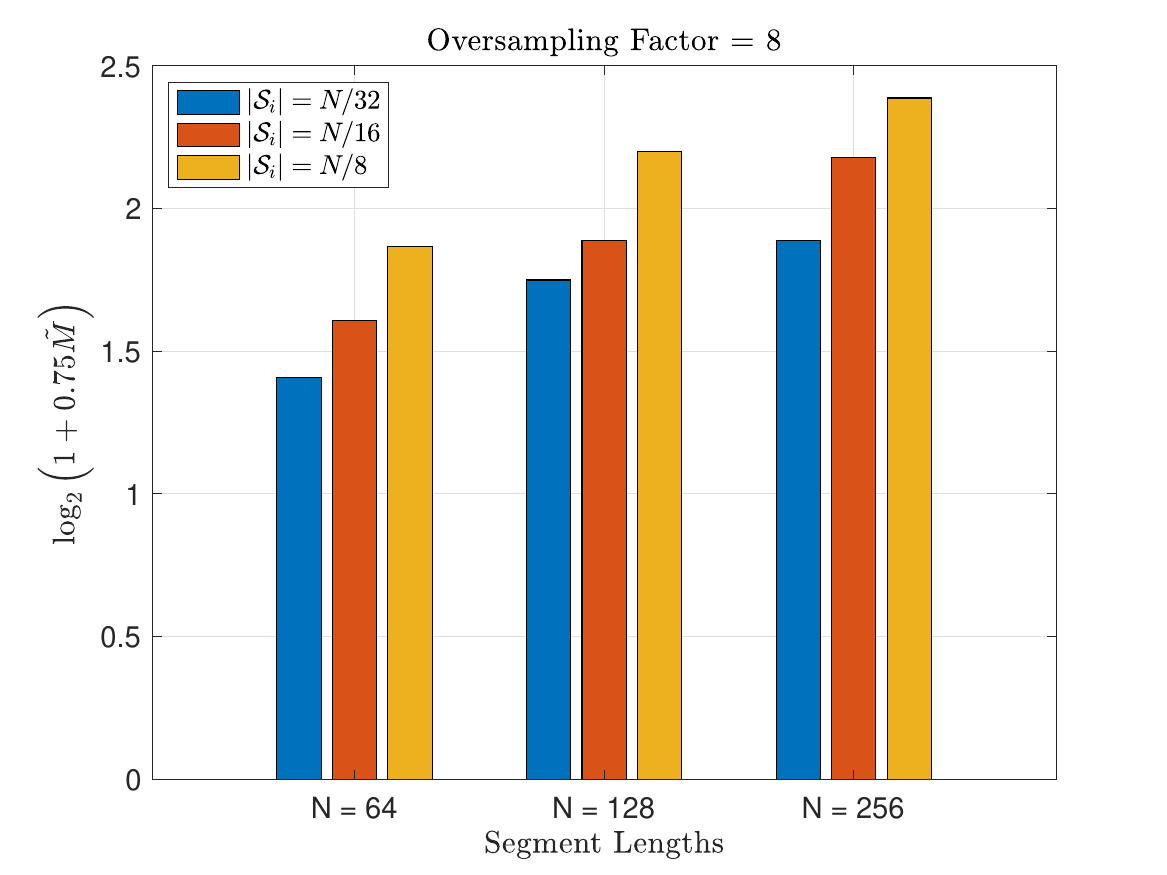}
    \caption{Empirical evaluation of $\log_2\left(1+0.75M\right)$ as a function of segment length ($N$) and $|\mathcal{S}_i|$ for $\mathrm{OF} = 8$. \textcolor{black}{The values of the segment length used in the numerical experiments are $N = 64$, $128$, and $256$. The values of $|\mathcal{S}_i|/N$ are set to $\frac{1}{8}$, $\frac{1}{16}$, and $\frac{1}{32}$.}}
    \label{fig:V_norm_OF8}
\end{figure}
\begin{figure}[t!]
    \hspace*{-.25cm}
    \includegraphics[scale = .49]{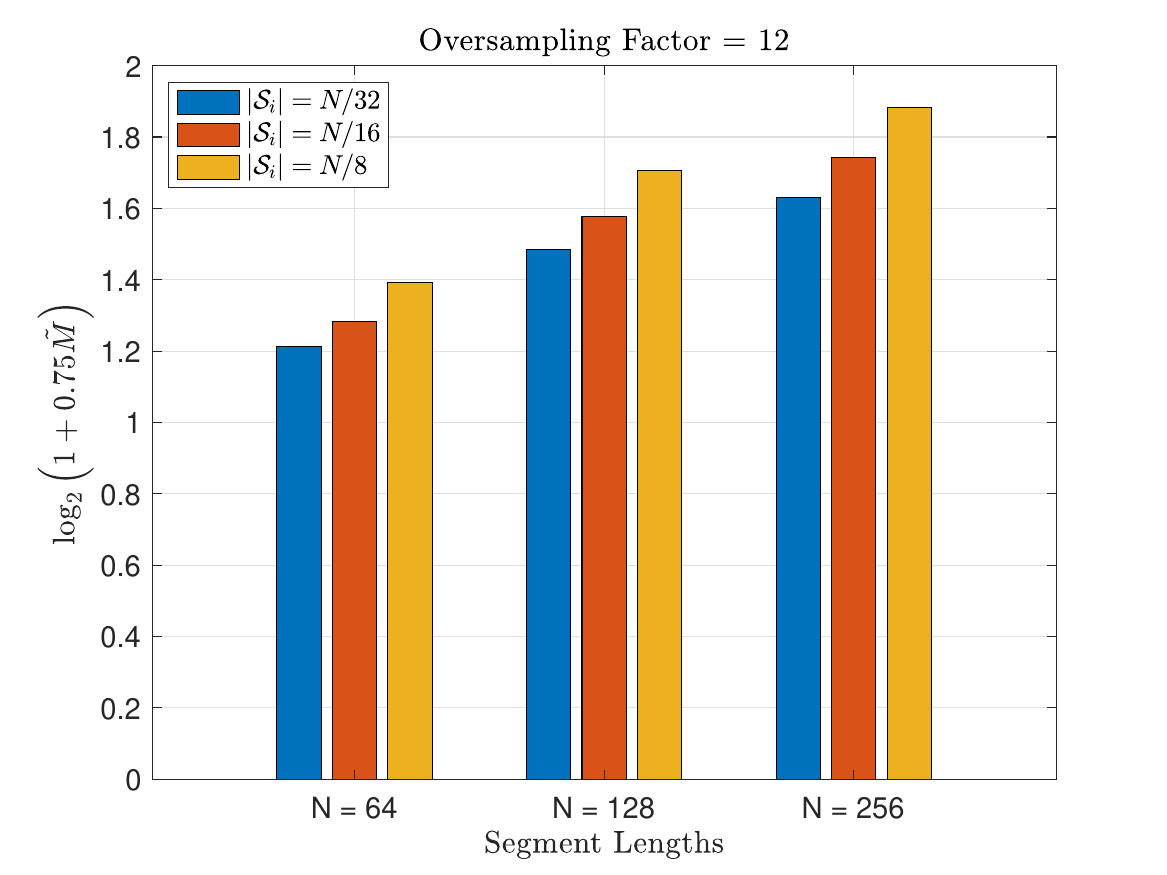}
    \caption{Empirical evaluation of $\log_2\left(1+0.75M\right)$ as a function of segment length ($N$) and $|\mathcal{S}_i|$ for $\mathrm{OF} = 12$. \textcolor{black}{The values of the segment length used in the numerical experiments are $N = 64$, $128$, and $256$. The values of $|\mathcal{S}_i|/N$ are set to $\frac{1}{8}$, $\frac{1}{16}$, and $\frac{1}{32}$.}}
    \label{fig:V_norm_OF12}
\end{figure}

Figures \ref{fig:V_norm_OF4}-\ref{fig:V_norm_OF12} depict the evaluation of $\log_2\left(1+0.75\tilde{M}\right)$ for $\mathrm{OF} =4,8$, and $12$, respectively. This term accounts for the extra bit resolution needed by the proposed algorithm (in addition to the 3 bits in the first term of equation \eqref{eq:b_req}) to satisfy the sufficient condition. Based on these evaluations, it can be seen that using a higher oversampling factor generally decreases the second term of equation \eqref{eq:b_req}. One insight that can be drawn from this observation is that a reduction in the modulo ADC resolution can be compensated by oversampling. Figures \ref{fig:V_norm_OF4}-\ref{fig:V_norm_OF12} also show that longer segment length\textcolor{black}{s} and frequent crossing\textcolor{black}{s} of the modulo thresholds increase $M$. 

To help interpret the observed trends, we recall that $M$ is the $\ell_{\infty}$-norm of the product of the Moore-Penrose inverse of $\mathbf{V}_{\mathcal{S}_i}$ and $\mathbf{V}_{\mathcal{S}_i^c}$. This quantity measures the maximum interference between the columns corresponding to folding indices and those outside the folding set. Therefore, the magnitude of $M$ is governed by two mechanisms: (i) the conditioning of $\mathbf{V}_{\mathcal{S}_i}$, which determines how much the pseudo-inverse operation amplifies the quantization noise, and (ii) the cross-coherence between $\mathbf{V}_{\mathcal{S}_i}$ and $\mathbf{V}_{\mathcal{S}_i^c}$. Increasing $\mathrm{OF}$ while keeping $|\mathcal{S}_i|$ fixed increases the dimension of the $|\mathcal{S}_i|$ column vectors of $\mathbf{V}_{\mathcal{S}_i}$. This leads to improved conditioning of $\mathbf{V}_{\mathcal{S}_{i}}$. In contrast, increasing $|\mathcal{S}_i|$ while keeping $\mathrm{OF}$ fixed has the opposite effect. The matrix $\mathbf{V}_{\mathcal{S}_i}$ becomes wider due to the increased number of column vectors, making $\mathbf{V}_{\mathcal{S}_i}$ more poorly conditioned. This results in a larger pseudo-inverse gain, which dominates the reduction in the number of columns of $\mathbf{V}_{\mathcal{S}_i^c}$.
%Increasing $\mathrm{OF}$ corresponds to an increase of the number of rows of $\mathbf{V}_{\mathcal{S}_i}$, thereby improving the conditioning of $\mathbf{V}_{\mathcal{S}_{i}}$. In contrast, increasing $|\mathcal{S}_i|$ increases the cross-coherence between $\mathbf{V}_{\mathcal{S}_i}$ and $\mathbf{V}_{\mathcal{S}_i^c}$, leading to larger values of $M$.}

As a final note in this numerical experiment, the sufficient condition for $b$ derived in the previous section may not be tight. In the succeeding numerical experiment, we show that the proposed algorithm can achieve the performance guarantees in Theorem \ref{theorem:MSE_guarantee}, even if we set $b = 4$. 

\subsection{Theoretical vs. Simulated MSE of the Proposed STFT-based Unfolding}\label{section:theo_vs_sim}
To demonstrate the validity of the theoretical MSE guarantees, we consider the input signal
\begin{align}\label{eq:test_sig}
    f(t) = \sum_{m = 1}^{150,000}A_m\cdot p_{\mathrm{rc}}(t - mT),
\end{align}
where $p_{\mathrm{rc}}(t)$ is a raised-cosine pulse with roll-off parameter $\beta = 0.25$. For the numerical simulation, we set the filter span of the raised-cosine pulse to 20. The peak amplitude of the raised-cosine pulse $p_{\mathrm{rc}}(t)$ is set to unity. The pulse amplitude $A_m$ is drawn uniformly from the interval $[-0.5,+1]$, and $T = 1$ second is the time difference between two consecutive pulses. Intuitively,signal $f(t)$ is constructed by generating 150,000 raised-cosine pulses with random amplitudes at every symbol period $T$. Consequently, the maximum frequency is $\frac{\omega_{m}}{2} = \frac{\pi(1+\beta)}{T}$. The interval is selected such that $f(t)$ has a non-zero mean value. We want to show that the algorithm can work on signals with a non-zero average value. This signal is fed to a low-resolution 4-bit modulo ADC with a sampling rate of $\frac{1}{T_{\mathrm{s}}} = \frac{\mathrm{OF}\omega_m}{2\pi}$ Hz to produce the quantized modulo samples $f_{\lambda',\mathrm{q}}[n]$. The dither sequence and modulo threshold are configured according to the modulo ADC parameters. 

For the STFT-based recovery algorithm, we set the roll-off parameter of the tapered cosine window to $\alpha = 0.50$ and window length $N = 64$. This choice of window parameters provides a balance between computational complexity and spectral leakage suppression. Note that the length of the tapered cosine window is also the segment length and the DFT size of the algorithm. To investigate the impact of spectral leakage, we consider two different spectral leakage values: $\delta_{\mathrm{SL}} = \frac{3\pi}{64},\;\frac{4\pi}{64}$ (resp. $K_{\mathrm{SL}} = 6,8$). Spectral leakage in the interval $\left(\rho\pi + \delta_{\mathrm{SL}},2\pi - \rho\pi - \delta_{\mathrm{SL}}\right)$ is assumed to be negligible. We will show later that this assumption is valid for this simulation setup since the theoretical and simulated MSE performances coincide. Equation \eqref{eq:lambda_p_req} is used to specify the modulo threshold $\lambda'$. The simulated performance is evaluated by taking the squared difference between the true samples $f[n]$ and the signal unfolding output $\hat{f}[n]$ and averaging the results over all $N_0$ samples. Note that the simulated MSE is computed using a single realization of $f(t)$. This is sufficient because, when the conditions on $\mathrm{OF}$ and $b$ in Theorem \ref{theorem:MSE_guarantee} are satisfied, there is no unfolding error and the MSE is due to the quantization noise, which forms an ergodic process. Consequently, the statistical average is well-approximated by the time average of a single realization of $f(t)$ for a sufficiently large $N_0$.

\begin{figure}[t!]
    \hspace*{-.51cm}
    \includegraphics[scale = .50]{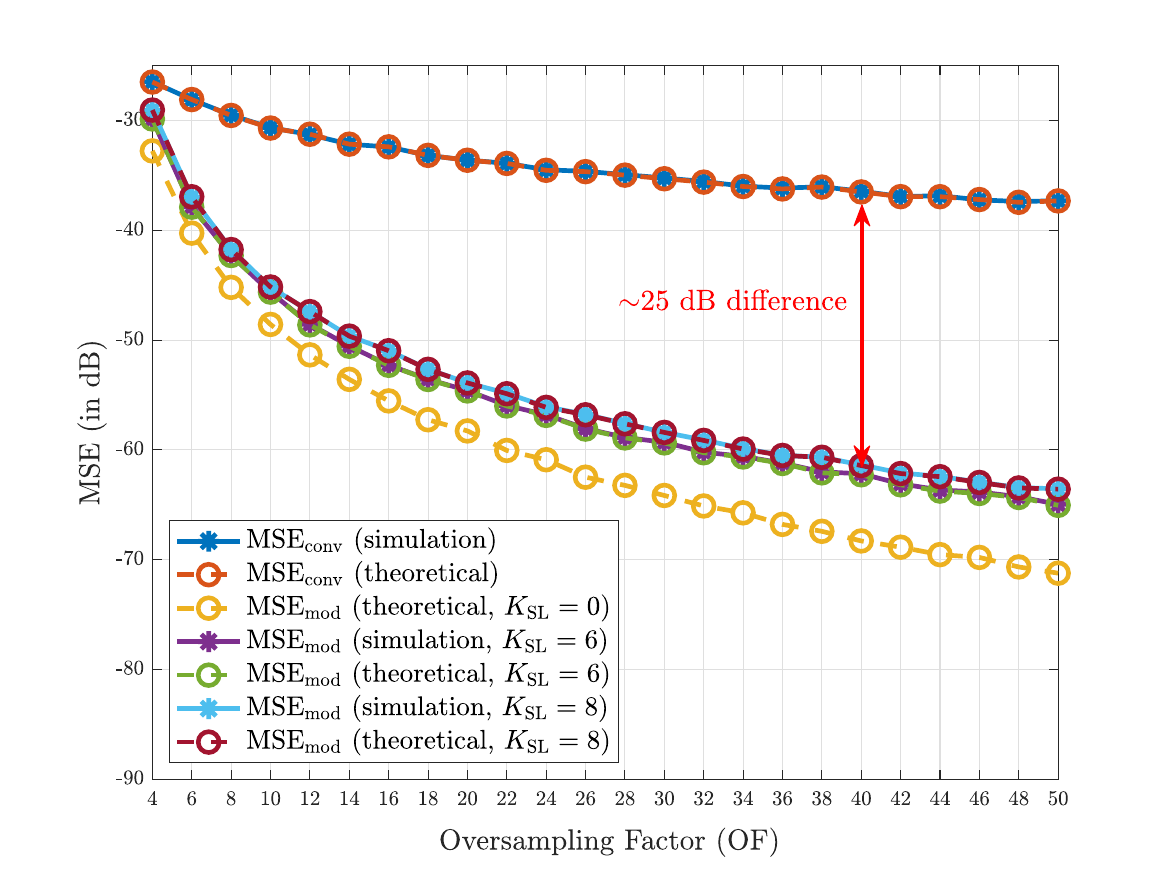}
    \caption{Simulated and theoretical MSE vs. oversampling factor of the proposed STFT-based unfolding algorithm for $N = 64$, $\alpha = 0.50$, $b = 4$ and $K_{\mathrm{SL}} = 6,8$. Superimposed in the figure are the simulated and theoretical MSE curves of a conventional ADC.}
    \label{fig:MSE_vs_OF}
\end{figure}

Figure \ref{fig:MSE_vs_OF} depicts the simulated MSE (in decibels) of the modulo ADC with STFT-based recovery and its theoretical MSE (based on Theorem \ref{theorem:MSE_guarantee}) under the aforementioned settings. The oversampling factor is swept from $\mathrm{OF} = 4$ to $\mathrm{OF} = 50$. The numerical results for \textcolor{black}{the} modulo ADC demonstrate that the derived MSE performance guarantee is accurate because it coincides with the simulated MSE for all $\mathrm{OF}$ and $K_{\mathrm{SL}}$ settings considered. It can be observed that larger $\delta_{\mathrm{SL}}$ increases the MSE of the STFT-based recovery method. This can be attributed to two factors: (a) the widening of the passband region of the digital LPF used in the last step of the unfolding method, and (b) the increase in $\lambda'$ according to equation \eqref{eq:lambda_p_req}. Consequently, a higher quantization noise power would be present at the digital LPF output.

%Consequently, more quantization noise are able to pass through the digital LPF. 

The theoretical MSE curve for $K_{\mathrm{SL}} = 0$ (i.e., spectral leakage neglected) is also superimposed in Figure \ref{fig:MSE_vs_OF}. It is evident from the figure that the slope of the $K_{\mathrm{SL}} = 0$ MSE curve is steeper in the high $\mathrm{OF}$ regime compared to those of $K_{\mathrm{SL}} = 6$ and $K_{\mathrm{SL}} = 8$ MSE curves. This validates the asymptotic MSE growth rates mentioned in Section \ref{section:performance_guarantee}. Nonetheless, the modulo ADC\textcolor{black}{, which uses our STFT-based recovery,} still outperforms the conventional ADC, even in the presence of spectral leakage. The gap between the MSE curves of the modulo ADC and conventional ADC widened as the $\mathrm{OF}$ is increased. At $\mathrm{OF} = 40$, the difference between the MSE incurred by the conventional ADC and that incurred by the proposed STFT-based recovery with $K_{\mathrm{SL}} = 8$ is approximately 25 dB. This result demonstrates that modulo ADCs can outperform conventional ADCs in oversampled settings, potentially enabling lower power consumption by relaxing resolution requirements.

\subsection{Comparison between the STFT-based recovery and HoD-based recovery}
We now compare the performance of the proposed STFT-based recovery method with an existing unfolding method for modulo sampling. Note that it is not practical to unfold $f(t)$ in \eqref{eq:test_sig} using the DFT-based methods described in \cite{Fourier-Prony,Shah:2023,Bernardo_ISIT2024,Azar:2022,Shah:2024,Bernardo2025TSP} because of its large $N_0$ and its lack of periodicity. Moreover, the thresholding-based recovery method in \cite[Algorithm 1]{Florescu:2022} is unsuitable for the given $f(t)$, because it requires a minimum separation between folding times to function correctly. The low-sampling rate variant \cite[Algorithm 2]{Florescu:2022} is likewise impractical because to its $\mathcal{O}(N_0^2)$ computational complexity. As a benchmark, we compared our proposed STFT-based recovery method with the HoD modulo recovery approach \cite{Bhandari:2017,Bhandari:2021}.

\begin{figure}[t!]
    \hspace*{-.51cm}
    \includegraphics[scale = .50]{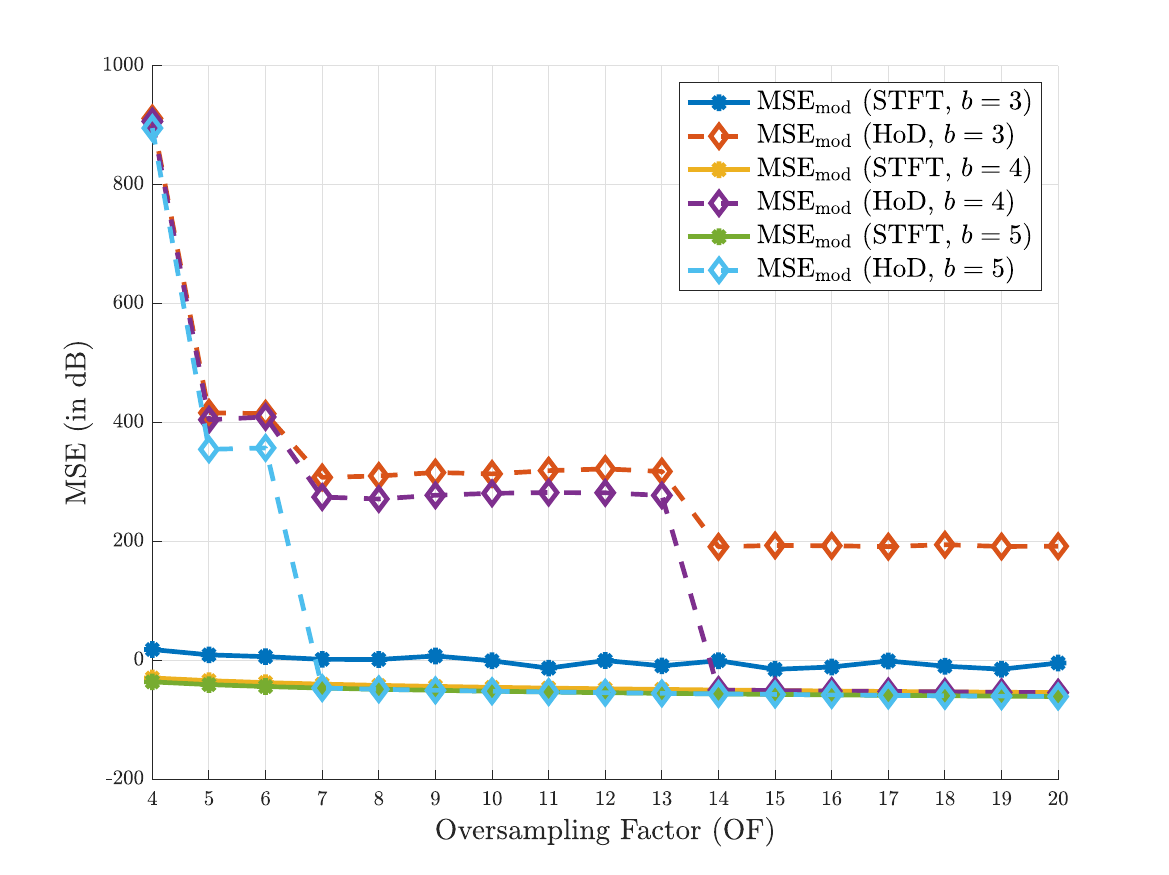}
    \caption{Simulated MSE vs. oversampling factor of the STFT-based recovery for $b = 3,4,5$. \textcolor{black}{Superimposed in the figure are the simulated MSE curves of the HoD-based recovery.}}
    \label{fig:slidingDFT_vs_HoD}
\end{figure}

Figure \ref{fig:slidingDFT_vs_HoD} shows the simulated MSE performance of the proposed STFT-based recovery method compared with the HoD-based recovery method for different values of $\mathrm{OF}$ and $b$. The parameters for the STFT-based method are the same as those in Section \ref{section:theo_vs_sim}, except for $\delta_{\mathrm{SL}} = \frac{3\pi}{64}$. For both methods, the modulo threshold $\lambda'$ is set according to \eqref{eq:lambda_p_req}, and the test signal $f(t)$ is given by \eqref{eq:test_sig}. When $b = 4$ or $b = 5$, the proposed method successfully unfolded the modulo ADC output across all $\mathrm{OF}$ values considered. In contrast, the HoD-based recovery shows very high MSE values for $\mathrm{OF} < 14$ when $b = 4$, and for $\mathrm{OF} < 7$ when $b = 5$. These results highlight the advantage of the proposed STFT-based recovery method over the HoD-based recovery in the low-sampling rate, low-resolution regime. The performance advantage of our proposed STFT-based recovery method over the HoD-based recovery method could be attributed to the availability of 1-bit folding information.

We also note that both methods fail to unfold the modulo ADC output when $b=3$, suggesting that $b>3$ may be a necessary condition for the successful operation of STFT-based recovery in its current form. Further research is needed to confirm this hypothesis.  Alternatively, employing other quantizer architectures in conjunction with modulo sampling may enable STFT-based unfolding at resolutions below $b=4$. For instance, the one-bit unlimited sampling approach in \cite{Graf:2019,Vaclav:2025} uses a 1-bit sigma–delta quantizer together with modulo sampling to implement a 1-bit modulo ADC.

\section{Summary and Future Directions}
\label{section:conclussion}

In this work, we considered a modulo ADC system in which the output samples are associated with a 1-bit folding information signal $c[n]$. Unfolded samples were obtained using the STFT-based recovery scheme described in Section \ref{subsection:signal_recovery}. The advantage of our proposed recovery scheme over existing Fourier-based recovery algorithms for modulo sampling is the significant reduction in the observation window length. This reduction in observation time \textcolor{black}{reduces} the computational complexity of the algorithm. We also provided sufficient conditions for the oversampling factor and ADC resolution to guarantee a certain MSE performance. When these sufficient conditions are met, the MSE performance of the modulo ADC that uses our proposed recovery method is better \textcolor{black}{than} that of the conventional ADCs. Moreover, we demonstrated that spectral leakage affects the asymptotic MSE growth rate of our STFT-based recovery method. These theoretical results are substantiated by the numerical experiments conducted in Section \ref{section:numerical_results}. Under the same modulo threshold setting and $b = 4,5$, the proposed STFT-based recovery method outperforms the HoD-based recovery method at low oversampling factor values. These findings help connect classical sampling theory with modulo quantization frameworks and point to new directions for designing low-resolution front ends with established performance guarantees.

It is worth mentioning that while the proposed unfolding scheme relies on the availability of a 1-bit folding information signal $c[n]$, the proposed recovery can be modified to work without $c[n]$. This can be achieved by performing robust sparse recovery of $\underline{\tilde{z}^{(i)}_{w}}[n]$ using an STFT-based version of LASSO-$B^2R^2$ \cite{Shah:2023} or an STFT-based version of OMP-based unfolding in \cite{Bernardo2025TSP}. This idea, together with its performance guarantees, is considered a potential subject for future research. We also note that the sufficient conditions derived for the oversampling factor and ADC resolution were not tight. Hence, another future direction is to tighten these conditions.

It is important to note that the performance guarantees presented in this study assume ideal modulo nonlinearities, that is, instantaneous and perfectly accurate folding without hysteresis or folding transients. This assumption represents a limitation of the current analysis, because real hardware implementations may deviate from this ideal behavior. Thus, we also intend to apply STFT-based recovery to generalized modulo sampling models and analyze how modulo hysteresis and folding transients affect the performance of the algorithm. Another direction is experimental validation of the proposed method under realistic analog impairments. Practical ADCs experience thermal noise, front-end nonlinearities, temperature drift, and other impairments that are not captured by the present theoretical model. Finally, it is of interest to apply this recovery algorithm to specific low-resolution signal processing systems (e.g., communication receivers and radars).

\begin{appendices}
\section{Proof of Proposition \ref{proposition:oversampling_req_gen}}\label{proof:appendix_A}

A necessary condition for $\mathbf{V}_{\mathcal{S}_i}$ to be full column rank is $|\mathcal{S}_{i}|\leq K$. Since the columns of $\mathbf{V}_{\mathcal{S}_i}$ are derived from the columns of the Fourier basis, they are linearly independent. Thus, the condition $|\mathcal{S}|\leq K$ is also sufficient. The number of discrete frequencies in the out-of-band region can be lower bounded by
\begin{align}\label{eq:K_LB}
   % K =& N - 2\left\lfloor\frac{\rho N}{2}\right\rfloor - K_{\mathrm{SL}}\nonumber\\
    K \geq& N(1-\rho) - K_{\mathrm{SL}}.
\end{align}
To establish $|\mathcal{S}_{i}|\leq K$, it suffices to show that \[|\mathcal{S}_i| \leq N(1-\rho) - K_{\mathrm{SL}}\] or, equivalently,
\begin{align}
    \mathrm{OF} \geq \frac{N}{N - |\mathcal{S}_{i}|-K_{\mathrm{SL}}}.
\end{align}
for all $i$. Here, we used $\mathrm{OF} = \frac{1}{\rho}$. The proof is completed by taking the maximum over all $|\mathcal{S}_i|$.

\section{Proof of Proposition 
\ref{proposition:oversampling_req}}\label{proof:appendix_B}

We first bound the size of $\mathcal{S}_i$ using an expression that is independent of the segment index. To do this, we note that a real bandlimited signal $f(t)$ crosses a fixed level $\ell$ in a time interval of length $T$ at most $2f_{B}T$ times, where $f_{\mathrm{B}}$ is the bandwidth of $f(t)$. Consider the time interval that corresponds to the $i$-th segment. Since $2f_{\mathrm{B}} = \frac{\omega_{m}}{2\pi}$ and $T = N T_{\mathrm{s}} = \frac{2\pi N}{\mathrm{OF}\cdot \omega_{m}}$, the number of times that $f(t)$ crosses level $\ell$ is at most $\frac{N}{\mathrm{OF}}$. Moreover, the number of $(2\mathbb{Z}+1)\lambda'$ folding levels in $f(t)$ is $2+ 2\lfloor\frac{\|f(t)\|_{\infty} - \lambda'}{2\lambda'}\rfloor$ as shown in \cite{Shah:2023}. Consequently, $|\mathcal{S}_i|$ is bounded as follows:
\begin{align}\label{eq:S_card_bound1}
    |\mathcal{S}_i|\;\leq& \;\frac{N}{\mathrm{OF}}\left(2+ 2\bigg\lfloor\frac{\|f(t)\|_{\infty} - \lambda'}{2\lambda'}\bigg\rfloor\right)\nonumber\\
    \;\leq&\; \frac{N}{\mathrm{OF}}\left(1+ \frac{\|f(t)\|_{\infty}}{\lambda'}\right)
\end{align}
for all $i$. The first line follows from multiplying the number of level crossings in a single level and the number of folding levels. The second line comes from the trivial \textcolor{black}{upper} bound of the floor function, i.e., $\lfloor x \rfloor \leq x$. Since the bound in equation \eqref{eq:S_card_bound1} is independent of $i$, this bound is also an upper bound for $\underset{i}{\max}\;|\mathcal{S}_i|$. Using equation \eqref{eq:S_card_bound1}, the following inequality holds:
\begin{align}\label{eq:OF_bound_general_tmp}
        \frac{N}{N - \frac{N}{\mathrm{OF}}\left(1+ \frac{\|f(t)\|_{\infty}}{\lambda'}\right) - K_{\mathrm{SL}}} \geq \frac{N}{N - \underset{i}{\max}\;|\mathcal{S}_i| - K_{\mathrm{SL}}}.
\end{align} 
Consequently, setting
\begin{align}
    \mathrm{OF} \geq \frac{N}{N - \frac{N}{\mathrm{OF}}\left(1+ \frac{\|f(t)\|_{\infty}}{\lambda'}\right)- K_{\mathrm{SL}}}
\end{align}
ensures that Proposition \ref{proposition:oversampling_req_gen} is satisfied by transitivity. We can isolate $\mathrm{OF}$ on one side to get
\begin{align}\label{eq:OF_result1}
    \mathrm{OF}\geq \frac{\frac{\|f(t)\|_{\infty}}{\lambda'} + 2}{1-\frac{K_{\mathrm{SL}}}{N}}.
\end{align}
Using \[\lambda' = \frac{\|f(t)\|_{\infty}}{\mathrm{OF}\cdot\left[1-\frac{K_{\mathrm{SL}}}{N}\right] - 2}\] as the modulo threshold satisfies the inequality in \eqref{eq:OF_result1}. 
However, since $\lambda' \leq \|f(t)\|_{\infty}$, \textcolor{black}{we get}
\begin{align}
    \mathrm{OF}\cdot\left[1-\frac{K_{\mathrm{SL}}}{N}\right] - 2 \geq 1.
    %\geq 1\;\;\Longrightarrow\;\;\mathrm{OF}\geq \frac{3}{1-\frac{K_{\mathrm{SL}}}{N}}
\end{align}
This implies that $\mathrm{OF}\geq \frac{3}{1-\frac{K_{\mathrm{SL}}}{N}}$ must also hold. The $\mathrm{OF}$ requirements are satisfied simultaneously if equation \eqref{eq:OF_requirement} holds.

\section{Proof of Theorem 
\ref{theorem:MSE_guarantee}}\label{proof:appendix_C}

As an initial step, an $\ell_{\infty}$-norm bound on the difference between the modulo residue pre-estimate $\mathbf{\tilde{z}}$ and modulo residue $\mathbf{z}$ is established:
\begingroup
\allowdisplaybreaks
 \begin{align}
    \|\underline{\mathbf{\tilde{z}}} - \underline{\mathbf{z}}\|_{\infty} =& \;\|\underline{\mathbf{\tilde{z}}_{\mathcal{S}_i}} - \underline{\mathbf{z}_{\mathcal{S}_i}}\|_{\infty}\nonumber\\
    = &\;\|\mathbf{V}_{\mathcal{S}_{i}}^{\dagger}\hat{\mathbf{F}}_{\mathrm{OOB}} - \underline{\mathbf{z}_{\mathcal{S}_i}}\|_{\infty}\nonumber\\
    = &\;\|\mathbf{V}_{\mathcal{S}_{i}}^{\dagger}\mathbf{V}_{\mathcal{S}_{i}}(\underline{\mathbf{z}_{\mathcal{S}_i}} + \underline{\boldsymbol{
    \epsilon
    }_{\mathcal{S}_i}}) +\mathbf{V}_{\mathcal{S}_{i}}^{\dagger}\mathbf{V}_{\mathcal{S}_{i}^c}\underline{\boldsymbol{
    \epsilon
    }_{\mathcal{S}_i^c}}  - \underline{\mathbf{z}_{\mathcal{S}_i}}\|_{\infty}\nonumber\\
    =&\;\|\mathbf{V}_{\mathcal{S}_{i}}^{\dagger}\mathbf{V}\underline{\boldsymbol{
    \epsilon
    }} \|_{\infty}
    \nonumber\\
    \leq &\;\left(1+\|\mathbf{V}_{\mathcal{S}_{i}}^{\dagger}\mathbf{V}_{\mathcal{S}^c_{i}} \|_{\infty}\right)\cdot\|\underline{\boldsymbol{
    \epsilon
    }}\|_{\infty}
    \nonumber\\
    \leq &\;\left(1+\|\mathbf{V}_{\mathcal{S}_{i}}^{\dagger}\mathbf{V}_{\mathcal{S}_i^c} \|_{\infty}\right)\cdot\left(\frac{6\lambda}{2^b}\right)\nonumber\\
    = &\;\left(1+\|\mathbf{V}_{\mathcal{S}_{i}}^{\dagger}\mathbf{V}_{\mathcal{S}_i^c}\|_{\infty}\right)\cdot\left(\frac{6\lambda'}{2^b-2}\right).
\end{align}
\endgroup
The first line comes from the fact that $\underline{\tilde{z}}^{(i)}[n] = \underline{{z}}^{(i)}[n] = 0$ for $n\notin \mathcal{S}_{i}$. The second and third lines come from equations \eqref{eq:F_OOB} and \eqref{eq:windowed_modulo_preestimate}. Because the last $\frac{\alpha N}{2}$ samples of $\tilde{z}_{w}^{(i-1)}[n]$ \textcolor{black}{are} added to the first $\frac{\alpha N}{2}$ samples of $\tilde{z}_{w}^{(i)}[n]$, the effect of the window function is removed. Consequently, the subscript $w$ is dropped in the second line. The fourth line holds because $\mathbf{V}_{\mathcal{S}_i}$ is full column rank by Proposition \ref{proposition:oversampling_req} and $\mathbf{V}_{\mathcal{S}_i}^\dagger\mathbf{V}_{\mathcal{S}_{i}}\underline{\boldsymbol{\epsilon}_{\mathcal{S}_i}} + \mathbf{V}_{\mathcal{S}_i}^\dagger\mathbf{V}_{\mathcal{S}_{i}^c}\underline{\boldsymbol{\epsilon}_{\mathcal{S}_i^c}} = \mathbf{V}_{\mathcal{S}_i}^\dagger\mathbf{V}\underline{\boldsymbol{\epsilon}}$. The fifth line follows from the property $\|\mathbf{A}\mathbf{x}\|_{\infty}\leq \|\mathbf{A}\|_{\infty}\|\mathbf{x}\|_{\infty}$. The sixth line holds because the triangle dither induces a quantization noise $\epsilon[n]$ whose range of amplitude is $\left(-\frac{3\lambda}{2^b},+\frac{3\lambda}{2^b}\right)$. Consequently, $\underline{\epsilon}[n]$ has amplitude in $\left(-\frac{6\lambda'}{2^b},+\frac{6\lambda'}{2^b}\right)$. Finally, the last line comes from the relationship between the modulo threshold $\lambda'$ and the ADC dynamic range parameter $\lambda$.

Suppose we let $M = \underset{i}{\max}\;\|\mathbf{V}_{\mathcal{S}_{i}}^{\dagger}\mathbf{V}_{\mathcal{S}_i^c}\|_{\infty}$. The first-order difference of the pre-estimate $\underline{\tilde{z}^{(i)}}[n]$ is at most $\frac{6\lambda'}{2^{b}-2}\left(1+M\right)$ away from the first-order difference of the true residue sample. Since a rounding operation to the nearest integer multiple of $2\lambda'$ is applied to the $\underline{\tilde{z}^{(i)}}[n]$ to get $\underline{\hat{z}^{(i)}}[n]$, perfect recovery of $z^{(i)}[n]$ after the application of the rounding operation and equation \eqref{eq:remove_firstorder_diff} is guaranteed if
\begin{align}
    \frac{6\lambda'}{2^{b}-2}(1+M) < \lambda',
\end{align}
or equivalently, \[3 + \log_2\left(1 + \frac{3}{4}M\right) < b.\]
This is satisfied by the assumption on $b$ stated in the theorem.

With perfect recovery of the $z^{(i)}[n]$, the terms $|z^{(i)}[n]-\hat{z}^{(i)}[n]| = 0$ for all $i$ and $n$. Thus, the MSE is solely due to the quantization noise power after filtering. From Section \ref{subsection:stat_quant_noise}, the quantization noise power is $\mathbb{E}\{|\epsilon[n]|^2\} = \frac{1}{4}(\frac{2\lambda}{2^b})^2$. This quantization noise power is spread evenly over the entire bandwidth ($-\pi$,$\pi$) since $\epsilon[n]$ is a white process. Considering the filtering operation via a digital LPF with passband region $\left(-\frac{\pi}{\mathrm{OF}} - \delta_{\mathrm{SL}},+\frac{\pi}{\mathrm{OF}} + \delta_{\mathrm{SL}}\right)$ after the unfolding step, the power of the filtered quantization noise is only $\frac{1}{\mathrm{OF}} + \frac{\delta_{\mathrm{SL}}}{\pi}$ of $\mathbb{E}\{|\epsilon^{(i)}[n]|^2\}$:
\begingroup
\allowdisplaybreaks
\begin{align}
    \mathbb{E}\left\{\big|\epsilon^{(i)}_{\mathrm{LPF}}[n]\big|^2\right\} =& \frac{\mathbb{E}\{|\epsilon^{(i)}[n]|^2\} }{\mathrm{OF}}\left(1+\frac{\delta_{\mathrm{SL}}\cdot\mathrm{OF}}{\pi}\right)\nonumber\\
    =& \frac{\lambda^2}{\mathrm{OF}\cdot2^{2b}}\left(1+\frac{\delta_{\mathrm{SL}}\cdot\mathrm{OF}}{\pi}\right)\nonumber\\
    =& \frac{(\lambda')^2}{\mathrm{OF}\cdot(2^{b}-2)^2}\left(1+\frac{\delta_{\mathrm{SL}}\cdot\mathrm{OF}}{\pi}\right).
\end{align}
\endgroup
The proof is completed by setting $\lambda' = \frac{\|f(t)\|_{\infty}}{\mathrm{OF}\cdot\left[1-\frac{K_{\mathrm{SL}}}{N}\right] - 2}$.

\end{appendices}

\renewcommand*{\bibfont}{\footnotesize}
\begingroup
\footnotesize  % or \small, \scriptsize, etc.
\printbibliography

@ARTICLE{Gray:1993,
  author={Gray, R.M. and Stockham, T.G.},
  journal={IEEE Transactions on Information Theory}, 
  title={Dithered quantizers}, 
  year={1993},
  volume={39},
  number={3},
  pages={805-812},
  doi={10.1109/18.256489}}

@misc{krishna,
      title={Unlimited Dynamic Range Analog-to-Digital Conversion}, 
      author={Adithya Krishna and Sunil Rudresh and Vishal Shaw and Hemanth Reddy Sabbella and Chandra Sekhar Seelamantula and Chetan Singh Thakur},
      year={2019},
      eprint={1911.09371},
      archivePrefix={arXiv},
      primaryClass={eess.SP},
      url={https://arxiv.org/abs/1911.09371}, 
}

@ARTICLE{UNO,
  author={Eamaz, Arian and Mishra, Kumar Vijay and Yeganegi, Farhang and Soltanalian, Mojtaba},
  journal={IEEE Transactions on Signal Processing}, 
  title={UNO: Unlimited Sampling Meets One-Bit Quantization}, 
  year={2024},
  volume={72},
  number={},
  pages={997-1014},
  doi={10.1109/TSP.2024.3356253}}

@INPROCEEDINGS{Shah:2023,
  author={Shah, Shaik Basheeruddin and Mulleti, Satish and Eldar, Yonina C.},
  booktitle={ICASSP 2023 - 2023 IEEE International Conference on Acoustics, Speech and Signal Processing (ICASSP)}, 
  title={Lasso-Based Fast Residual Recovery For Modulo Sampling}, 
  year={2023},
  volume={},
  number={},
  pages={1-5},
  doi={10.1109/ICASSP49357.2023.10097222}}

@INPROCEEDINGS{Feuillen:2023,
  author={Feuillen, Thomas and Shankar MRR, Bhavani and Bhandari, Ayush},
  booktitle={ICASSP 2023 - 2023 IEEE International Conference on Acoustics, Speech and Signal Processing (ICASSP)}, 
  title={Unlimited Sampling Radar: Life Below the Quantization Noise}, 
  year={2023},
  volume={},
  number={},
  pages={1-5},
  doi={10.1109/ICASSP49357.2023.10096015}}

@ARTICLE{Bhandari:2021,
  author={Bhandari, Ayush and Krahmer, Felix and Raskar, Ramesh},
  journal={IEEE Transactions on Signal Processing}, 
  title={On Unlimited Sampling and Reconstruction}, 
  year={2021},
  volume={69},
  number={},
  pages={3827-3839},
  doi={10.1109/TSP.2020.3041955}}

@article{Azar:2022,
  title={Robust unlimited sampling beyond modulo},
  author={Azar, Eyar and Mulleti, Satish and Eldar, Yonina C},
  journal={arXiv preprint arXiv:2206.14656},
  year={2022}
}

@ARTICLE{Romanov:2019,
  author={Romanov, Elad and Ordentlich, Or},
  journal={IEEE Signal Processing Letters}, 
  title={Above the Nyquist Rate, Modulo Folding Does Not Hurt}, 
  year={2019},
  volume={26},
  number={8},
  pages={1167-1171},
  doi={10.1109/LSP.2019.2923835}}

@book{Eldar, 
place={Cambridge},
title={Sampling Theory: Beyond Bandlimited Systems}, DOI={10.1017/CBO9780511762321}, publisher={Cambridge University Press}, 
author={Eldar, Y. C.}, year={2015}}

@ARTICLE{Fourier-Prony,
  author={Bhandari, A. and Krahmer, F. and Poskitt, T.},
journal={IEEE Trans. Signal Process.},
  title={Unlimited {Sampling} {From} {Theory} to {Practice}: {Fourier}-{Prony} {Recovery} and {Prototype} {ADC}}, 
  year={2022},
  volume={70},
  number={},
  pages={1131-1141},
  doi={10.1109/TSP.2021.3113497}}

@INPROCEEDINGS{Musa:2018,
  author={Musa, Osman and Jung, Peter and Goertz, Norbert},
  booktitle={2018 IEEE Global Conference on Signal and Information Processing (GlobalSIP)}, 
  title={Generalized Approximate Message Passing for Unlimited Sampling of Sparse Signals}, 
  year={2018},
  volume={},
  number={},
  pages={336-340},
  doi={10.1109/GlobalSIP.2018.8646332}}

@ARTICLE{Prasanna:2021,
  author={Prasanna, Dheeraj and Sriram, Chandrasekhar and Murthy, Chandra R.},
  journal={IEEE Signal Processing Letters}, 
  title={On the Identifiability of Sparse Vectors From Modulo Compressed Sensing Measurements}, 
  year={2021},
  volume={28},
  number={},
  pages={131-134},
  doi={10.1109/LSP.2020.3047584}}

@article{Shah:2021,
  title={Sparse signal recovery from modulo observations},
  author={Shah, Viraj and Hegde, Chinmay},
  journal={EURASIP Journal on Advances in Signal Processing},
  volume={2021},
  number={1},
  pages={1--17},
  year={2021},
  publisher={SpringerOpen}
}

@article{Mulleti:2023,
author = {Mulleti, Satish and Reznitskiy, Eliya and Savariego, Shlomi and Namer, Moshe and Glazer, Nimrod and Eldar, Yonina C.},
title = {A hardware prototype of wideband high-dynamic range analog-to-digital converter},
journal = {IET Circuits, Devices \& Systems},
volume = {17},
number = {4},
pages = {181-192},
doi = {https://doi.org/10.1049/cds2.12156},
url = {},
eprint = {https://ietresearch.onlinelibrary.wiley.com/doi/pdf/10.1049/cds2.12156},
year = {2023}
}

@INPROCEEDINGS{Bernardo_ISIT2024,
  author={Bernardo, Neil Irwin and Shah, Shaik Basheeruddin and Eldar, Yonina C.},
  booktitle={2024 IEEE International Symposium on Information Theory (ISIT)}, 
  title={Modulo Sampling with 1-Bit Side Information: Performance Guarantees in the Presence of Quantization}, 
  year={2024},
  volume={},
  number={},
  pages={3498-3503},
  doi={10.1109/ISIT57864.2024.10619324}}

@book{berger1971rate,
  title     = {Rate Distortion Theory: A Mathematical Basis for Data Compression},
  author    = {Toby Berger},
  year      = {1971},
  publisher = {Prentice-Hall},
  address   = {Englewood Cliffs, NJ, USA},
}

@ARTICLE{Shannon:1949,
  author={Shannon, C.E.},
  journal={Proceedings of the IRE}, 
  title={Communication in the Presence of Noise}, 
  year={1949},
  volume={37},
  number={1},
  pages={10-21},
  doi={10.1109/JRPROC.1949.232969}}

@ARTICLE{Walden:1999,
  author={Walden, R.H.},
  journal={IEEE Journal on Selected Areas in Communications}, 
  title={Analog-to-digital converter survey and analysis}, 
  year={1999},
  volume={17},
  number={4},
  pages={539-550},
  doi={10.1109/49.761034}}

@ARTICLE{Beckmann:2024,
  author={Beckmann, Matthias and Bhandari, Ayush and Iske, Meira},
  journal={IEEE Transactions on Computational Imaging}, 
  title={Fourier-Domain Inversion for the Modulo Radon Transform}, 
  year={2024},
  volume={10},
  number={},
  pages={653-665},
  doi={10.1109/TCI.2024.3388871}}

@INPROCEEDINGS{Guo:2023,
  author={Guo, Ruiming and Bhandari, Ayush},
  booktitle={ICASSP 2023 - 2023 IEEE International Conference on Acoustics, Speech and Signal Processing (ICASSP)}, 
  title={ITER-SIS: Robust Unlimited Sampling Via Iterative Signal Sieving}, 
  year={2023},
  volume={},
  number={},
  pages={1-5},
  doi={10.1109/ICASSP49357.2023.10094780}}

@INPROCEEDINGS{Shtendel:2024,
  author={Shtendel, Gal and Bhandari, Ayush},
  booktitle={ICASSP 2024 - 2024 IEEE International Conference on Acoustics, Speech and Signal Processing (ICASSP)}, 
  title={Dual-Channel Unlimited Sampling for Bandpass Signals}, 
  year={2024},
  volume={},
  number={},
  pages={9711-9715},
  doi={10.1109/ICASSP48485.2024.10446576}}

@ARTICLE{Ordentlich:2018,
  author={Ordentlich, Or and Tabak, Gizem and Hanumolu, Pavan Kumar and Singer, Andrew C. and Wornell, Gregory W.},
  journal={IEEE Journal of Selected Topics in Signal Processing}, 
  title={A Modulo-Based Architecture for Analog-to-Digital Conversion}, 
  year={2018},
  volume={12},
  number={5},
  pages={825-840},
  doi={10.1109/JSTSP.2018.2863189}}

@article{Rhee:2003,
author = {Rhee, Jehyuk and Joo, Youngjoong},
year = {2003},
month = {03},
pages = {360 - 361},
title = {Wide dynamic range CMOS image sensor with pixel level ADC},
volume = {39},
journal = {Electronics Letters},
doi = {10.1049/el:20030246}
}

@ARTICLE{Sasagawa:2016,
  author={Sasagawa, Kiyotaka and Yamaguchi, Takahiro and Haruta, Makito and Sunaga, Yoshinori and Takehara, Hironari and Takehara, Hiroaki and Noda, Toshihiko and Tokuda, Takashi and Ohta, Jun},
  journal={IEEE Transactions on Electron Devices}, 
  title={An Implantable CMOS Image Sensor With Self-Reset Pixels for Functional Brain Imaging}, 
  year={2016},
  volume={63},
  number={1},
  pages={215-222},
  doi={10.1109/TED.2015.2454435}}

@ARTICLE{Ordonez:2021,
  author={Ordoñez, Luis G. and Ferrand, Paul and Duarte, Melissa and Guillaud, Maxime and Yang, Ganghua},
  journal={IEEE Open Journal of the Communications Society}, 
  title={On Full-Duplex Radios With Modulo-ADCs}, 
  year={2021},
  volume={2},
  number={},
  pages={1279-1297},
  doi={10.1109/OJCOMS.2021.3085518}}

@INPROCEEDINGS{Bhandari:2020,
  author={Bhandari, Ayush and Krahmer, Felix},
  booktitle={2020 IEEE International Conference on Image Processing (ICIP)}, 
  title={HDR Imaging From Quantization Noise}, 
  year={2020},
  volume={},
  number={},
  pages={101-105},
  doi={10.1109/ICIP40778.2020.9190872}}

@ARTICLE{Liu:2023,
  author={Liu, Ziang and Bhandari, Ayush and Clerckx, Bruno},
  journal={IEEE Transactions on Communications}, 
  title={$\lambda$–MIMO: Massive MIMO Via Modulo Sampling}, 
  year={2023},
  volume={71},
  number={11},
  pages={6301-6315},
  doi={10.1109/TCOMM.2023.3305528}}

@ARTICLE{Zhang:2023,
  author={Zhang, Qi and Zhu, Jiang and Qu, Fengzhong and Soh, De Wen},
  journal={IEEE Transactions on Aerospace and Electronic Systems}, 
  title={Line Spectral Estimation via Unlimited Sampling}, 
  year={2024},
  volume={60},
  number={5},
  pages={7214-7231},
  doi={10.1109/TAES.2024.3413711}}

@INPROCEEDINGS{Geng:2023,
  author={Geng, Tianyu and Ji, Feng and Pratibha and Tay, Wee Peng},
  booktitle={ICASSP 2023 - 2023 IEEE International Conference on Acoustics, Speech and Signal Processing (ICASSP)}, 
  title={Modulo EEG Signal Recovery Using Transformer}, 
  year={2023},
  volume={},
  number={},
  pages={1-5},
  doi={10.1109/ICASSP49357.2023.10095357}}

@INPROCEEDINGS{Feuillen:2022,
  author={Feuillen, Thomas and Alaee-Kerahroodi, Mohammad and Bhandari, Ayush and R, Bhavani Shankar M. and Ottersten, Björn},
  booktitle={2022 IEEE Radar Conference (RadarConf22)}, 
  title={Unlimited Sampling for FMCW Radars: A Proof of Concept}, 
  year={2022},
  volume={},
  number={},
  pages={1-5},
  doi={10.1109/RadarConf2248738.2022.9764291}}

@ARTICLE{Shtendel:2023,
  author={Shtendel, Gal and Florescu, Dorian and Bhandari, Ayush},
  journal={IEEE Transactions on Signal Processing}, 
  title={Unlimited Sampling of Bandpass Signals: Computational Demodulation via Undersampling}, 
  year={2023},
  volume={71},
  number={},
  pages={4134-4145},
  doi={10.1109/TSP.2023.3314274}}

@INPROCEEDINGS{Laporte-Fauret:2018,
  author={Laporte-Fauret, Baptiste and Ferré, Guillaume and Dallet, Dominique and Minger, Bryce and Fuché, Loïc},
  booktitle={2018 25th IEEE International Conference on Electronics, Circuits and Systems (ICECS)}, 
  title={ADC Resolution for Simultaneous Reception of Two Signals with High Dynamic Range}, 
  year={2018},
  volume={},
  number={},
  pages={729-732},
  doi={10.1109/ICECS.2018.8617945}}

@online{MathWorksTukeyWindow,
  title = {Tukey Window},
  author = {{MathWorks}},
  year = {2024},
  url = {https://www.mathworks.com/help/signal/ref/tukeywin.html},
  note = {Accessed: 2024-9-23}
}

@article{Shah:2024,
      title={Compressed Sensing Based Residual Recovery Algorithms and Hardware for Modulo Sampling}, 
      author={Shaik Basheeruddin Shah and Satish Mulleti and Yonina C. Eldar},
      year={2024},
      journal={arXiv preprint arXiv:2412.12724},
      url={https://arxiv.org/abs/2412.12724}, 
}

@misc{Zhu:2024,
      title={A Modulo Sampling Hardware Prototype and Reconstruction Algorithm Evaluation}, 
      author={Jiang Zhu and Junnan Ma and Zhenlong Liu and Fengzhong Qu and Zheng Zhu and Qi Zhang},
      year={2024},
      eprint={2410.19383},
      archivePrefix={arXiv},
      primaryClass={eess.SP},
      url={https://arxiv.org/abs/2410.19383}, 
}

@misc{Zhang:2024,
      title={On the Identifiability from Modulo Measurements under DFT Sensing Matrix}, 
      author={Qi Zhang and Jiang Zhu and Fengzhong Qu and Zheng Zhu and De Wen Soh},
      year={2024},
      eprint={2401.00194},
      archivePrefix={arXiv},
      primaryClass={cs.IT},
      url={https://arxiv.org/abs/2401.00194}, 
}

@INPROCEEDINGS{Bhandari:2018,
  author={Bhandari, Ayush and Krahmer, Felix and Raskar, Ramesh},
  booktitle={2018 IEEE International Conference on Acoustics, Speech and Signal Processing (ICASSP)}, 
  title={Unlimited Sampling of Sparse Signals}, 
  year={2018},
  volume={},
  number={},
  pages={4569-4573},
  doi={10.1109/ICASSP.2018.8462231}}

@ARTICLE{Gong:2021,
  author={Gong, Yicheng and Gan, Lu and Liu, Hongqing},
  journal={IEEE Signal Processing Letters}, 
  title={Multi-Channel Modulo Samplers Constructed From Gaussian Integers}, 
  year={2021},
  volume={28},
  number={},
  pages={1828-1832},
  doi={10.1109/LSP.2021.3108526}}

@INPROCEEDINGS{Bhandari:2017,
  author={Bhandari, Ayush and Krahmer, Felix and Raskar, Ramesh},
  booktitle={2017 International Conference on Sampling Theory and Applications (SampTA)}, 
  title={On unlimited sampling}, 
  year={2017},
  volume={},
  number={},
  pages={31-35},
  doi={10.1109/SAMPTA.2017.8024471}}

@ARTICLE{Fernandez-Menduina:2022,
  author={Fernández-Menduiña, Samuel and Krahmer, Felix and Leus, Geert and Bhandari, Ayush},
  journal={IEEE Transactions on Signal Processing}, 
  title={Computational Array Signal Processing via Modulo Non-Linearities}, 
  year={2022},
  volume={70},
  number={},
  pages={2168-2179},
  doi={10.1109/TSP.2021.3101437}}

@INPROCEEDINGS{Gan:2020,
  author={GAN, Lu and Liu, Hongqing},
  booktitle={2020 IEEE 11th Sensor Array and Multichannel Signal Processing Workshop (SAM)}, 
  title={High Dynamic Range Sensing Using Multi-Channel Modulo Samplers}, 
  year={2020},
  volume={},
  number={},
  pages={1-5},
  doi={10.1109/SAM48682.2020.9104340}}

@misc{Zhu:2024b,
      title={Unleashing Dynamic Range and Resolution in Unlimited Sensing Framework via Novel Hardware}, 
      author={Yuliang Zhu and Ayush Bhandari},
      year={2024},
      eprint={2410.20193},
      archivePrefix={arXiv},
      primaryClass={eess.SP},
      url={https://arxiv.org/abs/2410.20193}, 
}

@ARTICLE{Guo:2024,
  author={Guo, Ruiming and Zhu, Yuliang and Bhandari, Ayush},
  journal={IEEE Transactions on Signal Processing}, 
  title={Sub-Nyquist USF Spectral Estimation: $K$ Frequencies With $6K+4$ Modulo Samples}, 
  year={2024},
  volume={72},
  number={},
  pages={5065-5076},
  doi={10.1109/TSP.2024.3469068}}

@INPROCEEDINGS{Florescu:2025,
  author={Florescu, Dorian},
  booktitle={ICASSP 2025 - 2025 IEEE International Conference on Acoustics, Speech and Signal Processing (ICASSP)}, 
  title={Multichannel Modulo Sampling with Unlimited Noise}, 
  year={2025},
  volume={},
  number={},
  pages={1-5},
  doi={10.1109/ICASSP49660.2025.10888122}}

@ARTICLE{Florescu:2022,
  author={Florescu, Dorian and Krahmer, Felix and Bhandari, Ayush},
  journal={IEEE Transactions on Signal Processing}, 
  title={The Surprising Benefits of Hysteresis in Unlimited Sampling: Theory, Algorithms and Experiments}, 
  year={2022},
  volume={70},
  number={},
  pages={616-630},
  doi={10.1109/TSP.2022.3142507}}

@INPROCEEDINGS{Geethu:2025,
  author={Joseph, Geethu},
  booktitle={ICASSP 2025 - 2025 IEEE International Conference on Acoustics, Speech and Signal Processing (ICASSP)}, 
  title={Noise-Resilient Unlimited Sampling and Recovery of Sparse Signals}, 
  year={2025},
  volume={},
  number={},
  pages={1-5},
  doi={10.1109/ICASSP49660.2025.10888741}}

@INPROCEEDINGS{Florescu:2022b,
  author={Florescu, Dorian and Bhandari, Ayush},
  booktitle={2022 IEEE International Symposium on Information Theory (ISIT)}, 
  title={Unlimited Sampling via Generalized Thresholding}, 
  year={2022},
  volume={},
  number={},
  pages={1606-1611},
  doi={10.1109/ISIT50566.2022.9834687}}

@INPROCEEDINGS{Florescu:2022c,
  author={Florescu, Dorian and Bhandari, Ayush},
  booktitle={ICASSP 2022 - 2022 IEEE International Conference on Acoustics, Speech and Signal Processing (ICASSP)}, 
  title={Unlimited Sampling with Local Averages}, 
  year={2022},
  volume={},
  number={},
  pages={5742-5746},
  doi={10.1109/ICASSP43922.2022.9747127}}

@INPROCEEDINGS{Florescu:2021,
  author={Florescu, Dorian and Krahmer, Felix and Bhandari, Ayush},
  booktitle={2021 55th Asilomar Conference on Signals, Systems, and Computers}, 
  title={Unlimited Sampling with Hysteresis}, 
  year={2021},
  volume={},
  number={},
  pages={831-835},
  doi={10.1109/IEEECONF53345.2021.9723306}}

@ARTICLE{Liu:2025,
  author={Liu, Ziang and Bhandari, Ayush and Clerckx, Bruno},
  journal={IEEE Communications Letters}, 
  title={Full-Duplex Beyond Self-Interference: The Unlimited Sensing Way}, 
  year={2025},
  volume={29},
  number={1},
  pages={165-169},
  doi={10.1109/LCOMM.2024.3505054}}

@INPROCEEDINGS{Zhang:2025,
      title={One-Bit-Aided Modulo Sampling for DOA Estimation}, 
      booktitle={2025 International Radar Conference},
      author={Qi Zhang and Jiang Zhu and Fengzhong Qu and De Wen Soh},
      year={2025},
}

@ARTICLE{Bernardo2025TSP,
  author={Bernardo, Neil Irwin and Shah, Shaik Basheeruddin and Eldar, Yonina C.},
  journal={IEEE Transactions on Signal Processing}, 
  title={Modulo Sampling: Performance Guarantees in The Presence of Quantization}, 
  year={2025},
  volume={},
  number={},
  pages={1-14},
  doi={10.1109/TSP.2025.3598420}}

@INPROCEEDINGS{Mulleti:2025,
  author={Mulleti, Satish and Appaiah, Kumar and Pillai, Sibi Raj B.},
  booktitle={ICASSP 2025 - 2025 IEEE International Conference on Acoustics, Speech and Signal Processing (ICASSP)}, 
  title={Low-Rate Modulo Folded ADC for Detecting Linearly Modulated Communication Symbols}, 
  year={2025},
  volume={},
  number={},
  pages={1-5},
  doi={10.1109/ICASSP49660.2025.10890267}}

@INPROCEEDINGS{Graf:2019,
  author={Graf, Olga and Bhandari, Ayush and Krahmer, Felix},
  booktitle={ICASSP 2019 - 2019 IEEE International Conference on Acoustics, Speech and Signal Processing (ICASSP)}, 
  title={One-bit Unlimited Sampling}, 
  year={2019},
  volume={},
  number={},
  pages={5102-5106},
  doi={10.1109/ICASSP.2019.8683266}}

@article{Itoh:1982,
author = {Kazuyoshi Itoh},
journal = {Appl. Opt.},
number = {14},
pages = {2470--2470},
publisher = {Optica Publishing Group},
title = {Analysis of the phase unwrapping algorithm},
volume = {21},
month = {Jul},
year = {1982},
url = {https://opg.optica.org/ao/abstract.cfm?URI=ao-21-14-2470},
doi = {10.1364/AO.21.002470},
}

@ARTICLE{Vaclav:2025,
  author={Pavl´ıcek, Vaclav and Bhandari, Ayush},
  journal={IEEE Journal of Selected Topics in Signal Processing}, 
  title={1-Bit Unlimited Sampling Beyond Fourier Domain: Low-Resolution Sampling of Quantization Noise}, 
  year={2025},
  volume={},
  number={},
  pages={1-12},
  doi={10.1109/JSTSP.2025.3603969}}

@INPROCEEDINGS{Yan:2025,
  author={Yan, Wenyi and Zhu, Ruixiang and Li, Zeyuan and Gan, Lu and Liu, Hongqing},
  booktitle={2025 International Conference on Sampling Theory and Applications (SampTA)}, 
  title={Parameter Selection in Complex-Valued Two-Channel Modulo ADC Sampling System}, 
  year={2025},
  volume={},
  number={},
  pages={1-5},
  doi={10.1109/SampTA64769.2025.11133533}}
\endgroup

\end{document}